\documentclass[journal]{IEEEtran}
\usepackage{amsmath,amsfonts}
\usepackage{algorithmic}
\usepackage{algorithm}
\usepackage{array}
\usepackage[caption=false,font=footnotesize,labelfont=rm,textfont=rm]{subfig}
\usepackage{textcomp}
\usepackage{stfloats}
\usepackage{url}
\usepackage{verbatim}
\usepackage{graphicx}
\usepackage{cite}
\usepackage{bm}
\usepackage{amssymb} 
\usepackage{mdwmath}
\usepackage{mdwtab}
\usepackage{eqparbox}
\usepackage{cuted}
\usepackage[hidelinks]{hyperref}

\newtheorem{proposition}{Proposition}

\newtheorem{theorem}{Theorem}

\begin{document}

\title{Rotatable Antenna Enabled Spectrum Sharing: Joint Antenna Orientation and Beamforming Design}

\author{Xingxiang Peng, Qingqing Wu, Ziyuan Zheng, Wen Chen, Yanze Zhu, and Ying Gao
\thanks{
    The authors are with the Department of Electronic Engineering, Shanghai Jiao Tong University, 200240, China 
    (e-mail: 
    \href{mailto:peng_xingxiang@sjtu.edu.cn}{\nolinkurl{peng_xingxiang@sjtu.edu.cn}};
    \href{mailto:qingqingwu@sjtu.edu.cn}{\nolinkurl{qingqingwu@sjtu.edu.cn}};
    \href{mailto:zhengziyuan2024@sjtu.edu.cn}{\nolinkurl{zhengziyuan2024@sjtu.edu.cn}};
    \href{mailto:wenchen@sjtu.edu.cn}{\nolinkurl{wenchen@sjtu.edu.cn}};
    \href{mailto:yanzezhu@sjtu.edu.cn}{\nolinkurl{yanzezhu@sjtu.edu.cn}};
    \href{mailto:yinggao@sjtu.edu.cn}{\nolinkurl{yinggao@sjtu.edu.cn}}).
    }
}




\maketitle

\begin{abstract}
    Conventional antenna arrays rely primarily on digital beamforming for spatial control. While adding more elements can narrow beamwidth and suppress interference, such scaling incurs prohibitive hardware and power costs. Rotatable antennas (RAs), which allow mechanical or electronic adjustment of element orientations, introduce a new degree of freedom to exploit spatial flexibility without enlarging the array. By dynamically optimizing orientations, RAs can substantially improve desired link alignment and interference suppression. This paper investigates RA-enabled multiple-input single-output (MISO) interference channels under co-channel spectrum sharing and formulates a weighted sum-rate maximization problem that jointly optimizes transmit beamforming and antenna orientations. To tackle this nonconvex problem, we develop an alternating optimization (AO) framework that integrates weighted minimum mean-square error (WMMSE)-based beamforming with Frank-Wolfe-based orientation updates. To reduce complexity, we further study orientation optimization under maximum-ratio transmission (MRT) and zero-forcing (ZF) beamforming schemes. For finite-resolution actuators, we construct spherical Fibonacci codebooks and design a cross-entropy method (CEM)-based algorithm for discrete orientation selection. Simulations show that integrating RAs with conventional beamforming markedly increases weighted sum-rate, with gains rising with element directivity. Under discrete orientation control, the proposed CEM algorithm consistently outperforms the nearest-projection baseline.
\end{abstract}


\section{Introduction}
\IEEEPARstart{T}{he} sixth-generation (6G) wireless networks aim to deliver ubiquitous connectivity under stringent requirements on data rate, latency, reliability, and energy efficiency \cite{intro1}. To support immersive extended reality, autonomous transportation, and massive IoT, 6G must markedly surpass 5G in spectral efficiency and user throughput while containing hardware and power costs, thereby motivating physical-layer technologies with additional spatial degrees of freedom, enhanced interference management, and agile adaptation to highly dynamic environments \cite{intro2,intro3,intro4}.

To meet the demand for substantial performance improvements in 6G, extremely large-scale multiple-input multiple-output (XL-MIMO) systems have been investigated to boost spatial degrees of freedom and enable near-field beam focusing with massive arrays \cite{intro5}. While XL-MIMO can deliver significant gains in capacity and interference suppression, it entails prohibitive hardware cost, high circuit power consumption, and considerable implementation complexity, hindering practical deployment \cite{intro6}. To alleviate these issues, intelligent reflecting surfaces (IRSs) have been proposed as a cost- and energy-efficient complement to existing infrastructure, enabling low-cost, low-power control of the radio environment \cite{intro7}. By tuning passive elements, IRSs reconfigure propagation to enhance desired link quality, suppress interference, and extend coverage \cite{intro8}. Explored use cases include interference management \cite{intro9,intro10}, cell-edge throughput enhancement \cite{intro11,intro12,intro121}, physical-layer security \cite{intro13,intro14}, and simultaneous wireless information and power transfer \cite{intro15,intro16}.

Following the paradigm of reconfiguring the wireless environment with IRS, a complementary research direction focuses on the underexplored potential of transceivers themselves. Rather than deploying additional reflective surfaces, this approach enhances the capability of existing antenna arrays by exploiting an often-overlooked degree of freedom: the position of each antenna element. This idea underpins movable antenna (MA) technology, where antenna positions can be dynamically optimized based on channel state information (CSI) \cite{it17,it18}. Compared with conventional fixed-position antennas, MAs can exploit local spatial channel variations within a confined region, even when the movement range is only a few wavelengths. By relocating elements to favorable points and reconfiguring the array geometry, they enhance desired-signal power and mitigate deep fading, place interferers in lower-gain regions for suppression, enable adaptive null steering and multibeam formation, and reshape the channel matrix to improve spatial multiplexing \cite{it19,it20,it21,it221}. These features have made MAs attractive in diverse application scenarios such as secure transmission \cite{it22}, integrated sensing and communication (ISAC) \cite{it23}, mobile edge computing (MEC) \cite{it24}, ultra-reliable low-latency communications (URLLC) \cite{it25}, and non-orthogonal multiple access (NOMA) systems \cite{it26}. In particular, \cite{it261} studies an MA-enabled MISO interference channel, in which antenna positions and transmit beamforming are jointly optimized to minimize total transmit power under per-user SINR constraints. Simulations show substantial power savings and improved spectrum reuse, with gains increasing with the size of the movable region. 

However, most existing MA investigations adopt isotropic element models, whereas practical base-station arrays typically employ downtilted directional elements. How to effectively harness antenna directivity within MA frameworks by incorporating realistic element radiation patterns and orientation control merits further investigation.

The recent concept of six-dimensional movable antennas (6DMA) generalizes conventional MA systems by enabling joint adaptation of position and orientation \cite{it27,it28,it29}. This added flexibility affords finer control of the effective aperture, improving link alignment, interference suppression, and spatial diversity exploitation \cite{it30,it31,it32}. As a lightweight, implementation-friendly variant, rotatable antennas (RAs) focus on the orientation dimension, a simplified yet impactful subset of 6DMA \cite{r1,it351,it33,it34,it35}. RAs introduce controllable angular adjustment into conventional arrays, providing additional spatial freedom without physical displacement. Compared with full MA systems that require continuous positional tracking and incur higher mechanical complexity, power consumption, and latency, RAs are easier to integrate, more energy-efficient, and compatible with existing planar arrays, making them attractive in scenarios where rapid adaptation is needed but mechanical movement is impractical. Beyond implementation benefits, RAs are particularly effective when practical antenna directivity is considered. For instance, in interference-limited regimes, steering each element’s boresight within a spherical cap leverages the directional radiation pattern to keep the intended user in the main lobe while pushing dominant interferers into lower-gain regions. This directivity-aware orientation enables precise beam alignment and strong interference suppression, whereas MAs primarily enhance channel rank and multipath diversity through positional changes. Thus, RA and MA offer complementary advantages: RAs provide low-overhead beamforming and interference management with directional elements, while MAs improve spatial multiplexing and diversity in suitable deployments.

Building on this background, this paper investigates RA-enabled MISO interference channels under co-channel spectrum sharing, where cross-link interference is the primary performance bottleneck. It is worth noting that conventional base stations typically employ downtilted directional antennas statically configured at installation, thereby fixing the radiation pattern and limiting adaptation to dynamic interference conditions. In scenarios such as spatially correlated interference, near–far asymmetry, or high user mobility, conventional digital or hybrid beamforming alone often struggles to suppress persistent cross-link interference due to fixed element directivity and tilt. In contrast, each transmit element in our proposed architecture is a directional antenna with a steerable boresight constrained to a spherical cap, enabling dynamic reshaping of the interference environment. This steering harnesses antenna directivity to focus energy toward intended users while suppressing dominant interferers, yielding an interference geometry that is more amenable to beamforming optimization and thereby increasing the weighted sum-rate in interference-limited scenarios. Overall, the main contributions of this work are summarized as follows:

\begin{itemize}
    \item To the best of our knowledge, this work presents the first study of RA-enabled MISO interference channels under co-channel spectrum sharing. We formulate a weighted sum-rate maximization problem that jointly optimizes transmit beamforming and antenna orientations, subject to per-transmitter power and spherical-cap orientation constraints. The problem is highly nonconvex because the objective couples beamforming vectors and orientation variables, and the spherical-cap constraints render the feasible set nonconvex as well.
    \item We develop an alternating optimization (AO) framework to tackle this nonconvex problem. Specifically, the beamforming vectors are updated using the weighted minimum mean-square error (WMMSE) method under per-transmitter power constraints. For the orientation variables, we propose a Frank–Wolfe-based algorithm operating on the spherical-cap manifold. This approach employs a closed-form linear oracle and Armijo backtracking to maintain feasibility and ensure monotonic improvement of the weighted sum-rate throughout the iterations.
    \item To promote practical deployment, we further introduce two extensions: (i) low-complexity solutions based on maximum-ratio transmission (MRT) and zero-forcing (ZF) beamforming schemes; and (ii) a discrete-orientation framework that uses spherical Fibonacci codebooks for nearly uniform sampling on the spherical cap, along with a cross-entropy method (CEM)-based algorithm for high-performance orientation selection.
    \item Simulations are conducted to validate the proposed algorithms under various practical setups. Results demonstrate that the integration of RA with conventional beamforming significantly improves the weighted sum-rate. Performance gains increase with both element directivity and array size. Even limited zenith angular freedom offers considerable gains, which eventually saturate at moderate zenith angles. For discrete actuation, the combination of a spherical Fibonacci codebook and CEM clearly and consistently outperforms the nearest-projection baseline.
\end{itemize}

The remainder of this paper is organized as follows. Section~II introduces the system model and formulates the weighted sum-rate maximization problem. Section~III develops an AO framework integrating WMMSE beamforming with Frank–Wolfe-based orientation updates. Section~IV presents low-complexity solutions under MRT and ZF beamforming schemes. Section~V discusses discrete orientation optimization using a CEM-based algorithm. Section~VI provides numerical results, and Section~VII concludes the paper.

\textit{Notations:} Bold lowercase, bold uppercase, and calligraphic letters denote vectors, matrices, and sets, respectively. The operators $(\cdot)^{T}$, $(\cdot)^{H}$, and $(\cdot)^{*}$ denote transpose, Hermitian transpose, and complex conjugate, respectively. The Euclidean norm of a vector $\bm{x}$ is denoted by $\|\bm{x}\|_{2}$, the modulus of a complex scalar $x$ by $|x|$, and the cardinality of a set $\mathcal{X}$ by $|\mathcal{X}|$. The inner product of vectors $\bm{x}$ and $\bm{y}$ is written as $\langle \bm{x},\bm{y}\rangle=\bm{x}^{H}\bm{y}$. The real-part operator is $\Re\{\cdot\}$, and $\bm{I}$ denotes the identity matrix of appropriate size. $[x]_{+}\triangleq \max\{0,x\}$. The natural and base-2 logarithms are denoted by $\ln(\cdot)$ and $\log_{2}(\cdot)$, respectively, while $e$ and $j\triangleq \sqrt{-1}$ denote Euler’s number and the imaginary unit. The distribution $\mathcal{CN}(0,\sigma^{2})$ denotes a circularly symmetric complex Gaussian random variable with zero mean and variance $\sigma^{2}$, and $\mathcal{U}[a,b]$ denotes the continuous uniform distribution over $[a,b]$. Expectation is denoted by $\mathbb{E}\{\cdot\}$, the floor operator by $\lfloor\cdot\rfloor$, and $\mathcal{O}(\cdot)$ denotes the computational complexity order. The indicator function $\mathbf{1}\{\cdot\}$ equals $1$ if the condition inside is true and $0$ otherwise.

\section{System Model and Problem Formulation}

\begin{figure}[t]
	\begin{center}
		\includegraphics[width=0.4\textwidth]{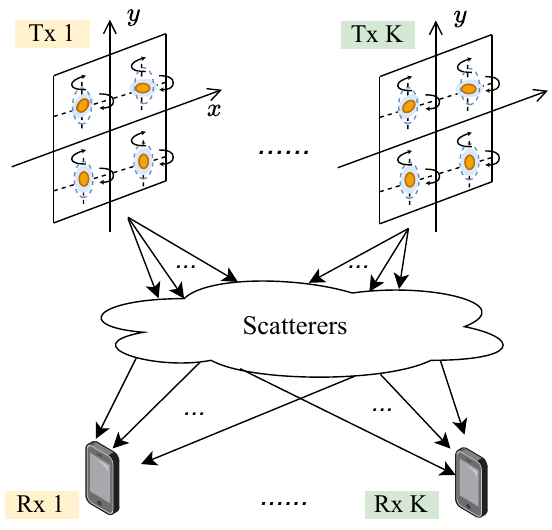}
		\caption{RA-enabled MISO interference channel, where each transmitter serves a dedicated user.}
		\label{fig:sysModel}
	\end{center}
\end{figure}

As illustrated in Fig.~\ref{fig:sysModel}, we consider a $K$-pair MISO interference channel where each transmitter is equipped with an $M_x \times M_y$ uniform planar array (UPA) of RAs to serve a single-antenna user\footnote{Users employ fixed omnidirectional antennas without rotation, which is practical for size/power-constrained devices. While we focus on the $K$-pair interference channel, the proposed framework can be extended to a multiuser downlink where one transmitter serves multiple users simultaneously.}. All transmitters share the same frequency band, resulting in an interference channel scenario among communication pairs. Without loss of generality, the UPAs are placed on the global $x$--$y$ plane with their default boresight pointing to the $+z$ direction. Let $\bm{t}_{k,0} \in \mathbb{R}^3$ denote the Cartesian coordinates of the center of transmitter $k$'s UPA, and let $d$ be the inter-element spacing. The elements are indexed by $(m_x,m_y)$ where $m_x \in \{1,2,\ldots,M_x\}$ and $m_y \in \{1,2,\ldots,M_y\}$. For notational convenience, we map the double index to a single index $m =  m_x + (m_y-1) M_x$, with $m \in \{1,2,\ldots,M\}$ and $M=M_xM_y$. The global coordinates of element $m$ at transmitter $k$ are then given by
\begin{align}
    \bm{t}_{k,m}
    = \bm{t}_{k,0}
    + d\!\begin{bmatrix}
    [m-1]_{M_x} - \tfrac{M_x-1}{2} \\
    \Big\lfloor \tfrac{m-1}{M_x} \Big\rfloor - \tfrac{M_y-1}{2} \\
    0
    \end{bmatrix},
\end{align}
where $[\cdot]_{M_x}$ denotes the modulo-$M_x$ operation.

\subsection{RA Model}

\begin{figure}[!t]
  \centering
  \subfloat[Directional power pattern]{%
    \includegraphics[width=0.42\linewidth]{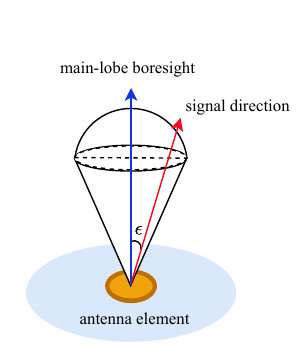}%
    \label{fig:dg}}
    \hfill
    \subfloat[Rotational angles]{%
    \includegraphics[width=0.42\linewidth]{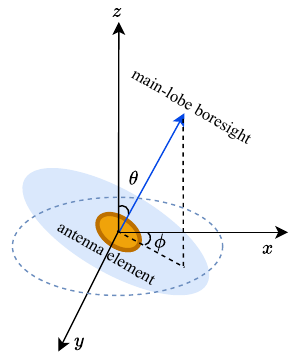}%
    \label{fig:ram}}
  \caption{Directional power pattern and rotation-angle parameterization of RA elements.}
  \label{fig:ramdfd}
\end{figure}

\begin{figure}[t]
	\begin{center}
		\includegraphics[width=0.42\textwidth]{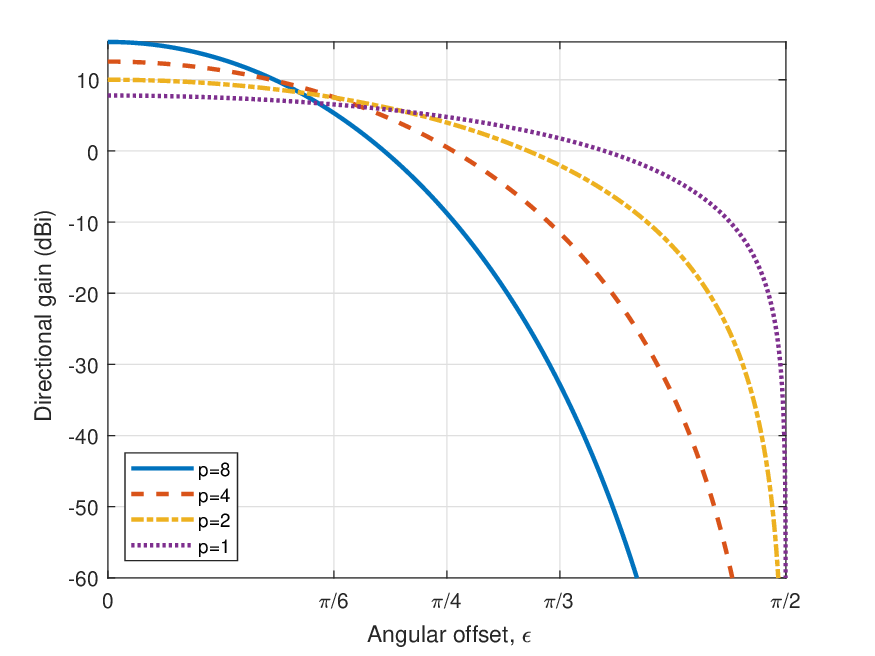}
		\caption{Directional gain $G(\epsilon)$ for different $p$, illustrating narrower main-lobes and higher boresight concentration as $p$ increases.}
		\label{Digain}
	\end{center}
\end{figure}

In practical deployments, base-station arrays commonly utilize downtilted directional elements that are statically configured during installation, resulting in a fixed coverage pattern. As traffic and mobility evolve, this inflexibility may lead to beam misalignment, fluctuating signal quality, and reduced network capacity. In contrast, RAs with rotatable boresights enable dynamic beam alignment, directional nulling, and real-time coverage shaping, improving spectral efficiency and link reliability.  Accordingly, we model each RA element as a directional radiator featuring a steerable boresight.

\subsubsection{Implementation of RAs}

RAs can be realized either mechanically or electronically~\cite{it351}. In mechanical designs, each element is mounted on a gimbal or servo drive that physically steers the boresight to the desired orientation; this offers fine angular resolution and a wide rotation range at the expense of actuation latency and potential mechanical wear. In electronic designs, reconfigurable feeding networks or tunable metasurfaces reshape the aperture current distribution to emulate rotation, enabling rapid and repeatable orientation reconfiguration without moving parts and with higher reliability. Both approaches provide directional boresight control with distinct trade-offs in accuracy, agility, complexity, and cost.

\subsubsection{Directional Power Pattern of RAs}

The directional gain of each RA element depends on the angular offset $\epsilon$ between the signal direction and the element's main-lobe boresight, as illustrated in Fig.~\ref{fig:ramdfd}(a), and can be modeled by a cosine pattern \cite[Chap.~2]{nr1}:
\begin{align}
    G(\epsilon) =
    \begin{cases}
        \kappa_{\text{max}} \cos^{2p}(\epsilon), & \epsilon \in \left[0, \tfrac{\pi}{2}\right], \\[4pt]
        0, & \text{otherwise},
    \end{cases}
    \label{eq:RA_pattern}
\end{align}
where $\kappa_{\text{max}}=2(2p+1)$ denotes the maximum gain achieved when the signal is aligned with the boresight, and $p$ is the directivity factor controlling the main-lobe width\footnote{Note that the coefficient $\kappa_{\text{max}}=2(2p+1)$ follows from directivity normalization over the full sphere, i.e., $\int_{4\pi} G(\epsilon)\,\text{d}\Omega=4\pi$.}. For illustration, Fig.~\ref{Digain} plots the resulting gain patterns for several values of $p$, showing that increasing $p$ leads to a more directive pattern with a narrower main-lobe and greater energy concentration toward the main-lobe boresight.

\subsubsection{Rotational Angle Parameterization of RAs}

As depicted in Fig.~\ref{fig:ramdfd}(b), the orientation of the main-lobe boresight for the $m$-th element at transmitter $k$ is parameterized by a zenith angle $\theta_{k,m} \in [0, \tfrac{\pi}{2}]$ (measured from the positive $z$-axis) and an azimuth angle $\phi_{k,m} \in [-\pi, \pi)$ (measured from the $x$-axis in the $x$–$y$ plane, counterclockwise). The corresponding boresight unit vector is then given by:
\begin{align}
    \label{ffunc}
    \bm{f}_{k,m}(\theta_{k,m},\phi_{k,m}) = 
    \begin{bmatrix}
    \sin\theta_{k,m} \cos\phi_{k,m} \\
    \sin\theta_{k,m} \sin\phi_{k,m} \\
    \cos\theta_{k,m}
    \end{bmatrix},
\end{align}
which satisfies $\|\bm{f}_{k,m}\|_{2} = 1$ by construction. To incorporate mechanical constraints and mitigate mutual coupling between adjacent elements, the zenith angle is subject to \cite{r1}
\begin{align}
    \label{fconstr}
    0 \le \theta_{k,m} \le \theta_{\max}, \quad \forall k,m,
\end{align}
where $\theta_{\max} \in [0,\tfrac{\pi}{2}]$. This angular constraint on $\theta_{k,m}$ is equivalent to the following constraint on the boresight vector
\begin{align}
\cos(\theta_{\max}) \le \bm{f}_{k,m}^{T}\bm{e}_z \le 1, \qquad \forall k,m,
\end{align}
where $\bm{e}_z = [0,0,1]^T$ is the unit vector along the $z$-axis.

\subsection{Channel Model}
We consider a narrowband far-field scenario and employ a channel model that incorporates the directional gain of the RA elements by using the Friis transmission formula to account for both distance and orientation.

\subsubsection{Line-of-Sight (LoS) channel component}
Let $\bm u_n\in\mathbb{R}^3$ denote the global coordinates of user $n$. Based on the RA power pattern in \eqref{eq:RA_pattern} and the Friis transmission equation \cite[Chap.~2]{nr1}, the link power gain between element $m$ of transmitter $k$ and user $n$ is given by
\begin{align}\label{cgm}
    G_{k,m,n}^{\text{LoS}}(\bm{f}_{k,m}) &= \beta_0r_{k,m,n}^{-2} G(\epsilon_{k,m,n}) \notag \\
    &=  \beta_0r_{k,m,n}^{-2}\kappa_{\max} 
      \Big[\frac{\bm{f}_{k,m}^{\!T}(\bm{u}_n-\bm{t}_{k,m})}{r_{k,m,n}}\Big]_{+}^{2p},
\end{align}
where $\beta_0 = \left(\frac{\lambda}{4\pi }\right)^2$ is the free-space reference gain constant, $\lambda$ denotes the wavelength, $r_{k,m,n} = \|\bm{t}_{k,m} - \bm{u}_n\|_2$ is the distance between element $m$ of transmitter $k$ and user $n$, and $\frac{\bm{f}_{k,m}^T (\bm{u}_n - \bm{t}_{k,m})}{r_{k,m,n}} = \cos(\epsilon_{k,m,n})$ represents the cosine of the angle between the boresight $\bm{f}_{k,m}$ and the LoS direction to user $n$, as illustrated in Fig.~\ref{fig:ramdfd}(a). The LoS channel coefficient is the square root of the link power gain with a propagation phase, which can be expressed as
\begin{align}
    \label{loschae}
    h_{k,m,n}^{\text{LoS}}(\bm{f}_{k,m}) 
    = \sqrt{G_{k,m,n}^{\text{LoS}}(\bm{f}_{k,m})}\,
      e^{-j\frac{2\pi}{\lambda}r_{k,m,n}}.
\end{align}
Thus, the channel coefficient captures both the directional antenna gain and the propagation-induced phase shift.

\subsubsection{Non-Line-of-Sight (NLoS) channel component}
We also consider a scattering environment with $Q$ spatially distributed clusters, located at $\{\bm{s}_q\in\mathbb{R}^3\}_{q=1}^Q$. Following the same principle as in \eqref{cgm}, the link power gain between element $m$ of transmitter $k$ and cluster $q$ is
\begin{align}
    {G}_{k,m,q}(\bm{f}_{k,m}) & = \beta_0\tilde{r}_{k,m,q}^{-2} G(\epsilon_{k,m,q}) \notag \\ 
    &= \beta_0\tilde{r}_{k,m,q}^{-2} \kappa_{\max} 
      \Big[\frac{\bm{f}_{k,m}^{\!T}(\bm{s}_q-\bm{t}_{k,m})}{\tilde{r}_{k,m,q}}\Big]_{+}^{\!2p},
\end{align}
where $\tilde{r}_{k,m,q} = \|\bm{t}_{k,m} - \bm{s}_q\|_2$ is the element-to-cluster distance, and $\frac{\bm{f}_{k,m}^{\!T}(\bm{s}_q-\bm{t}_{k,m})}{\tilde{r}_{k,m,q}} = \cos(\epsilon_{k,m,q})$ denotes the cosine of the angle between the boresight and the direction to cluster $q$.  
Under a bi-static scattering model \cite[Chap.~2]{nr1}, the NLoS channel coefficient from element $m$ of transmitter $k$ to user $n$ is then given by
\begin{align}
    \label{nloschae}
    &{h}_{k,m,n}^{\text{NLoS}}(\bm{f}_{k,m}) \notag \\
    &= \sum_{q=1}^Q \sqrt{{G}_{k,m,q}(\bm{f}_{k,m})}\cdot\sqrt{\frac{\sigma_q}{4\pi\hat{r}_{q,n}^{2}}} \cdot e^{-j\tfrac{2\pi}{\lambda}(\tilde{r}_{k,m,q}+\hat{r}_{q,n})+j\chi_q}  \notag \\
    &= \sum_{q=1}^Q \sqrt{\frac{\sigma_q\, {G}_{k,m,q}(\bm{f}_{k,m})}{4\pi\hat{r}_{q,n}^2}}
      e^{-j\tfrac{2\pi}{\lambda}(\tilde{r}_{k,m,q}+\hat{r}_{q,n})+j\chi_q},
\end{align}
where $\hat{r}_{q,n} = \|\bm{s}_q - \bm{u}_n \|_2$ is the cluster-to-user distance, $\sigma_q$ denotes the radar cross section (RCS) of cluster $q$, and $\chi_q$ is a random phase uniformly distributed over $[0, 2\pi)$. The factor $\sqrt{\sigma_q/(4\pi \hat{r}_{q,n}^{2})}$ arises from the bi-static scattering model, in which the re-radiated field has amplitude coefficient $\sqrt{\sigma_q/(4\pi)}$ and decays as $1/\hat r_{q,n}$ over the second hop \cite{nr1}.

\subsubsection{Overall channel representation}
By superimposing the LoS and NLoS components, the overall multipath channel between transmitter $k$ and user $n$ can be expressed as
\begin{align} \label{overallChan}
    \bm{h}_{k,n} (\bm{F}_{k})
    = \bm{h}_{k,n}^{\text{LoS}}(\bm{F}_{k}) + \bm{h}_{k,n}^{\text{NLoS}}(\bm{F}_{k}),
\end{align}
where $\bm{F}_{k} = [\bm{f}_{k,1},\ldots,\bm{f}_{k,M}]$, $\bm{h}_{k,n}^{\text{LoS}}(\bm{F}_{k}) = [h_{k,1,n}^{\text{LoS}}(\bm{f}_{k,1}),\ldots,h_{k,M,n}^{\text{LoS}}(\bm{f}_{k,M}) ]^{\!T}$, and 
$\bm{h}_{k,n}^{\text{NLoS}}(\bm{F}_{k}) = [h_{k,1,n}^{\text{NLoS}}(\bm{f}_{k,1}),\ldots,h_{k,M,n}^{\text{NLoS}}(\bm{f}_{k,M}) ]^{\!T}$. 
It follows that variations in $\{\bm{F}_k\}_{k=1}^{K}$ rotate the RA directional patterns, thereby altering the multipath channel in \eqref{overallChan}.

\subsection{Signal Model}
For each transmission pair $k$, let $\bm{w}_k\in\mathbb{C}^{M}$ denote the beamforming vector and $s_k$ the information symbol with $\mathbb{E}[|s_k|^2]=1$. The received signal at user $k$ is then given by
\begin{align}
    y_k \;=\; \bm{h}_{k,k}^{H}\,\bm{w}_{k}\,s_k
    \;+\; \sum_{n\ne k}^{K} \bm{h}_{n,k}^{H}\,\bm{w}_{n}\,s_n
    \;+\; z_k,
\end{align}
where $z_k\sim\mathcal{CN}(0,\delta_k^2)$ denotes complex additive white Gaussian noise (AWGN) at user $k$. Accordingly, the achievable rate for each user $k$ is given by
\begin{align}\label{eq:SINR}
    R_k
    = \log_{2}\!\left(1+\frac{\big|\bm{h}_{k,k}^{H}\,\bm{w}_{k}\big|^2}
    {\sum\limits_{n\neq k}^{K} \big|\bm{h}_{n,k}^{H}\bm{w}_{n}\big|^2 + \delta_k^2}\right).
\end{align}

\subsection{Problem Formulation}
Our goal is to jointly optimize the beamforming vectors and the element orientations to enhance the network spectral efficiency. Let $\{\alpha_k\geq 0 \}_{k=1}^K$ denote the weighting factors associated with the $K$ transmission pairs. The weighted sum-rate maximization problem is formulated as
\begin{align}
    \max_{\{\bm{w}_k,\bm{F}_k\}_{k=1}^K}
    \quad & \sum_{k=1}^{K} \alpha_k R_k \label{p1}\\
    \text{s.t.}\qquad 
    & \|\bm{w}_k\|_2^2 \le P_{k,\max}, \quad \forall k, \label{eq:pk}\\
    & \cos(\theta_{\max}) \le \bm{f}_{k,m}^{\!T}\bm{e}_z \le 1, \quad \forall k,m, \label{eq:orient_param}\\
    & \|\bm{f}_{k,m}\|_2 = 1,\quad \forall k,m, \label{eq:unitnormd}
\end{align}
where \eqref{eq:pk} imposes a per-transmitter power constraint, and \eqref{eq:orient_param}–\eqref{eq:unitnormd} enforce the mechanical rotation bounds and unit-norm boresight vectors.  Problem~\eqref{p1} is nonconvex and challenging because the objective is quadratic in the beamformers and nonlinear in the orientation variables through the orientation-dependent channels, yielding a nonlinear, nonseparable coupling. Moreover, the spherical-cap constraints \eqref{eq:orient_param}–\eqref{eq:unitnormd} render the feasible set nonconvex.

\section{The AO-Based Solution}

To handle the strong coupling between transmit beamforming and antenna orientations in Problem~\eqref{p1}, we propose an AO-based framework that alternates between WMMSE beamforming and Frank–Wolfe orientation updates. Starting from a feasible point, the framework iteratively alternates between two steps: (i) a beamforming step that applies the WMMSE method to update the beamformers with fixed orientations, and (ii) an orientation step that applies the Frank–Wolfe-based algorithm with Armijo backtracking to update the orientations with fixed beamformers. Each subproblem is feasible and guarantees a non-decreasing objective value, thereby ensuring monotonic convergence to a stationary point of Problem~\eqref{p1}.

\subsection{Optimization for Beamforming Vectors}
With fixed antenna orientations, Problem~\eqref{p1} reduces to 
\begin{align}
    \max_{\{\bm{w}_k\}_{k=1}^K}
    \quad & \sum_{k=1}^{K} \alpha_k R_k \label{p2}\\
    \text{s.t.}\qquad 
    & \|\bm{w}_k\|_2^2 \le P_{k,\max}, \,\, \forall k, \label{eq:p1k}
\end{align}
which can be readily solved by the WMMSE method \cite{r2}. Specifically, Problem~\eqref{p2} can be equivalently transformed into the following WMMSE minimization:
\begin{align}
    \min_{\{\bm{w}_k\},\,\{u_k\},\,\{v_k\}}
    \quad & \sum_{k=1}^{K} \alpha_k \big( v_k\, e_k - \log v_k \big) \label{eq:wmmse}\\
    \text{s.t.}\quad\qquad 
    & \|\bm{w}_k\|_2^2 \le P_{k,\max},\,\, \forall k, 
\end{align}
where $e_k \;=\; \mathbb{E}\!\big[\,|u_k^{*} y_k - s_k|^2\,\big] 
= \big|u_k^{*}\bm{h}_{k,k}^{H}\bm{w}_{k}-1\big|^2 
+ \sum_{n\neq k}^{K} \big|u_k^{*}\bm{h}_{n,k}^{H}\bm{w}_{n}\big|^2 
+ |u_k|^2\,\delta_k^2$. Note that $e_k$ is the MSE of the recovered signal at user $k$, and $u_k$ and $v_k$ denote the scalar equalizer and nonnegative weight, respectively. Problem~\eqref{eq:wmmse} is block-wise convex and admits closed-form updates in an alternating manner:

(1) Receive filter update: For fixed $\{\bm{w}_k\}$,
\begin{align}
    u_k^{\star}
    = \frac{\bm{h}_{k,k}^{H}\bm{w}_k}{\sum_{n=1}^{K}\big|\bm{h}_{n,k}^{H}\bm{w}_{n}\big|^2 + \delta_k^2},
    \quad \forall k. \label{eq:u_update}
\end{align}

(2) Weight update: For fixed $\{\bm{w}_k\}$ and $\{u_k\}$,
\begin{align}
    v_k^{\star}= e_k^{-1},\quad \forall k. \label{eq:v_update}
\end{align}

(3) Beamformer update: For fixed $\{u_k\}$ and $\{v_k\}$, the optimal solution of each $\bm{w}_k$ can be derived as
\begin{align}
    \bm{w}_k^{\star}
    &= \big(\sum_{i=1}^{K} \alpha_i v_i\,|u_i|^2\, \bm{h}_{k,i}\bm{h}_{k,i}^{H} + \mu_k \bm{I}\big)^{-1} (\alpha_k v_k\, u_k^{*}\, \bm{h}_{k,k}), \,\,  \forall k. \label{eq:w_update}
\end{align}
where $\mu_k\ge 0$ is the Lagrange multiplier chosen (e.g., via bisection method) to satisfy $\|\bm{w}_k^{\star}\|_2^2 = P_{k,\max}$ if $\mu_k>0$.

\subsection{Optimization for Antenna Orientations}
With fixed beamforming vectors $\{\bm w_k\}_{k=1}^K$, the optimization with respect to (w.r.t.) antenna orientations is given by:
\begin{align}
    \max_{\{\bm{F}_k\}_{k=1}^K} \quad & 
    \sum_{k=1}^{K} \alpha_k 
    \log_{2}\!\left(1+\frac{|\bm{h}_{k,k}^{H}\bm{w}_{k}|^2}
    {\sum_{n\neq k}^{K} |\bm{h}_{n,k}^{H}\bm{w}_{n}|^2 + \delta_k^2}\right)
    \label{eq:orientation_optimization} \\
    \text{s.t.} \quad \quad & 
    \cos(\theta_{\max}) \le \bm{f}_{k,m}^T\bm{e}_z \le 1, \quad \forall k,m,
    \label{eq:zenith_constraint} \\
    & \|\bm{f}_{k,m}\|_2 = 1, \quad \forall k,m.
    \label{eq:unit_norm_constraint}
\end{align}
Problem~\eqref{eq:orientation_optimization} is nonconvex for two reasons: (i) the objective couples all elements through the channel responses $\{\bm h_{n,k}\}$, and (ii) the spherical-cap constraints \eqref{eq:zenith_constraint}–\eqref{eq:unit_norm_constraint} induce a compact but nonconvex feasible region. In what follows, we develop a Frank–Wolfe-based algorithm to tackle this problem by tailoring the scheme to the spherical-cap constraint. While the classical Frank–Wolfe method is designed for compact convex sets, our approach combines tangent-space linearization with Armijo backtracking and yields monotonic improvement (as shown later in Theorem~\ref{thm:fw_convergence}).

To proceed, for each antenna element $(k,m)$ we define the per-element spherical cap on the unit sphere
\begin{equation}
    \mathcal{C}_{\text{cap}}
    \triangleq
    \big\{\bm x\in\mathbb{R}^3:\ \|\bm x\|_2=1,\ \bm x^{T}\bm e_z \ge \cos(\theta_{\max})\big\},
    \label{eq:single_cap}
\end{equation}
and consequently the overall feasible set of Problem~\eqref{eq:orientation_optimization}  is the Cartesian product of these caps,
\begin{equation}
    \mathcal{C}
    \triangleq
    \prod_{k=1}^{K}\prod_{m=1}^{M}\mathcal{C}_{\text{cap}}.
    \label{spcap}
\end{equation}
Next, we incorporate this structure into a Frank–Wolfe procedure. At iteration $t$, given the current orientations $\{\bm f^{(t)}_{k,m}\}$, we compute the blockwise gradients of the weighted sum-rate w.r.t. the antenna orientations (explicit expressions are derived in Appendix~\ref{apdix1})
\begin{align}
    \bm g^{(t)}_{k,m} \triangleq \nabla_{\bm f_{k,m}} \left(
    \sum_{\ell=1}^K \alpha_\ell R_\ell(\{\bm F_k\})\right),\,\, \forall k,m.
\end{align}
We then project each gradient onto the tangent space via $\bm{T}^{(t)}_{k,m}=\bm{I}-\bm f^{(t)}_{k,m}(\bm f^{(t)}_{k,m})^{\!T}$. Linearizing the objective around $\{\bm f^{(t)}_{k,m}\}$ yields the separable oracle as follows
\begin{align}
    \label{eq:linear_oracle}
    \{\bm s^{(t)}_{k,m}\} \in 
    \arg\max_{\{\bm x_{k,m}\} \in \mathcal{C}} 
    \sum_{k,m} \left\langle \bm{T}^{(t)}_{k,m}\bm g^{(t)}_{k,m}, \bm x_{k,m} \right\rangle,
\end{align}
which decouples across elements; equivalently, each $(k,m)$ solves an independent maximization over the spherical cap $\mathcal{C}_{\text{cap}}$. The linear oracle in \eqref{eq:linear_oracle} admits the following closed-form solution: if $\bm{T}^{(t)}_{k,m}\bm{g}^{(t)}_{k,m} = 0$, set $\bm{s}^{(t)}_{k,m} = \bm{f}^{(t)}_{k,m}$; otherwise, the solution is given by \eqref{eq:closed_form_solution}, shown at the bottom of the page, 
\begin{figure*}[b] 
	\setlength{\arraycolsep}{5pt}
	\hrulefill
	\vspace*{4pt}
	\setlength{\arraycolsep}{0.0em}
    \begin{align}
        \bm s^{(t)}_{k,m} =
        \begin{cases}
            \widehat{\bm g}^{(t)}_{k,m}, & 
            (\widehat{\bm g}^{(t)}_{k,m})^T\bm e_z \ge  \cos(\theta_{\max}), \\[4pt]
            \left[\sqrt{1 -  \cos^2(\theta_{\max})} \,
            \dfrac{\widehat{\bm g}^{(t)}_{k,m,xy}}
            {\|\widehat{\bm g}^{(t)}_{k,m,xy}\|_2},\ \cos(\theta_{\max}) \right]^{T}, & 
            (\widehat{\bm g}^{(t)}_{k,m})^T\bm e_z < \cos(\theta_{\max}),\ \|\widehat{\bm g}^{(t)}_{k,m,xy}\|_2>0, \\[8pt]
            \left[\sqrt{1-\cos^2(\theta_{\max})}\, \bm u,\ \cos(\theta_{\max}) \right]^{T}, &
            (\widehat{\bm g}^{(t)}_{k,m})^T\bm e_z < \cos(\theta_{\max}),\ \|\widehat{\bm g}^{(t)}_{k,m,xy}\|_2=0,
        \end{cases}
        \label{eq:closed_form_solution}
    \end{align}
\end{figure*}
where $\widehat{\bm g}^{(t)}_{k,m}=\bm{T}^{(t)}_{k,m}\bm{g}^{(t)}_{k,m}/\|\bm{T}^{(t)}_{k,m}\bm{g}^{(t)}_{k,m}\|_2$ is the unit-normalized gradient, $\widehat{\bm g}^{(t)}_{k,m,xy}\in\mathbb{R}^2$ denotes the $x$–$y$ components (i.e., the $xy$-plane projection) of $\widehat{\bm g}^{(t)}_{k,m}$, and $\bm u\in\mathbb{R}^2$ is any unit vector. The Frank-Wolfe-based search directions are given by
\begin{equation}
    \bm d^{(t)}_{k,m} \triangleq \bm s^{(t)}_{k,m} - \bm f^{(t)}_{k,m}, \,\, \forall k,m,
\end{equation}
with which the antenna orientations are updated as
\begin{equation}
    \bm f^{(t+1)}_{k,m}
    \;=\;
     \frac{\bm f^{(t)}_{k,m} + \rho^{(t)} \bm d^{(t)}_{k,m}}
             {\big\|\bm f^{(t)}_{k,m} + \rho^{(t)} \bm d^{(t)}_{k,m}\big\|_2}, \,\,\forall k,m,
    \label{eq:update_rule}
\end{equation}
where $\rho^{(t)} \in (0,1]$ is the stepsize. Note that, since both the current point $\bm f^{(t)}_{k,m}$ and the reference point $\bm s^{(t)}_{k,m}$ lie in the feasible spherical cap, their convex combination, i.e., $\bm f^{(t)}_{k,m}+\rho^{(t)}\bm d^{(t)}_{k,m}$, automatically satisfies the zenith-angle constraint~\eqref{eq:zenith_constraint}. The subsequent normalization enforces the unit-norm constraint~\eqref{eq:unit_norm_constraint}, thereby ensuring that the updated orientation $\bm f^{(t+1)}_{k,m}$ remains feasible. The stepsize $\rho^{(t)}$ is selected by Armijo backtracking. Specifically, the stepsize $\rho^{(t)}$ is accepted if the following Armijo sufficient-increase condition holds:
\begin{align}
    \label{eq:armijo_condition} 
    R\!\big(\{\bm f^{(t+1)}_{k,m}\}\big)
    \ \ge\
    R\!\big(\{\bm f^{(t)}_{k,m}\}\big)
    \;+\; c_{\mathrm A}\,\rho^{(t)}\,\sigma^{(t)},
\end{align}
where $R \triangleq \sum_{\ell=1}^K\alpha_\ell R_\ell$ denotes the objective value of Problem~\eqref{eq:orientation_optimization}, $c_{\text{A}} \in (0,1)$ is the Armijo parameter, and $\sigma^{(t)}$ represents the Frank-Wolfe gap defined as
\begin{align}
    \sigma^{(t)}=\sum_{k,m} \left\langle \bm T^{(t)}_{k,m}\bm g^{(t)}_{k,m},\, \bm d^{(t)}_{k,m}\right\rangle.
\end{align}
\begin{proposition}
    The Frank-Wolfe gap $\sigma^{(t)}$ is non-negative.
\end{proposition}
\begin{IEEEproof}
    Please refer to Appendix~\ref{apdix20}.
\end{IEEEproof}

\begin{theorem}
    \label{thm:fw_convergence}
    At each iteration $t$, there exists a stepsize $\rho^{(t)} > 0$ such that the sufficient-increase condition \eqref{eq:armijo_condition} holds.
\end{theorem}

\begin{IEEEproof}
    Please refer to Appendix~\ref{apdix2}.
\end{IEEEproof}

The Armijo condition \eqref{eq:armijo_condition} guarantees a sufficient increase at each accepted stepsize, and hence the objective sequence $R\!\big(\{\bm f^{(t)}_{k,m}\}\big)$ is monotonically nondecreasing. Moreover, since the feasible set is compact and the weighted sum-rate is bounded under the finite power budget, the sequence $R\!\big(\{\bm f^{(t)}_{k,m}\}\big)$ is bounded and therefore convergent. The Frank–Wolfe–based procedure for solving Problem \eqref{eq:orientation_optimization} is summarized in Algorithm~\ref{alg:fw_orientation}.

\begin{algorithm}[t]  \small
\caption{Frank-Wolfe-Based Algorithm for Antenna Orientation Optimization}
\label{alg:fw_orientation}
\begin{algorithmic}[1]
\REQUIRE Initial feasible orientations $\{\bm f^{(0)}_{k,m}\}$, tolerance $\epsilon > 0$, maximum iterations $T_{\max}$
\ENSURE Optimized orientations $\{\bm f^{(t)}_{k,m}\}$
\STATE $t \leftarrow 0$;
\STATE Compute $R^{(0)} \leftarrow R(\{\bm f^{(0)}_{k,m}\})$;
\WHILE{$t < T_{\max}$ and ($\frac{|R^{(t)} - R^{(t-1)}|}{|R^{(t-1)}|} > \epsilon$)}
    \FOR{each antenna element $(k,m)$}
        \STATE Compute gradient $\bm g^{(t)}_{k,m} \leftarrow  \nabla_{\bm f_{k,m}} \left(
    \sum_{\ell=1}^K \alpha_\ell R_\ell\right)$;
    \ENDFOR
    \FOR{each antenna element $(k,m)$}
        \STATE Solve oracle: $\bm s^{(t)}_{k,m} \leftarrow \arg\max_{\bm x \in \mathcal{C}_{\text{cap}}} \langle \bm T_{k,m}^{(t)} \bm g^{(t)}_{k,m}, \bm x \rangle$ via \eqref{eq:closed_form_solution};
        \STATE Compute direction: $\bm d^{(t)}_{k,m} \leftarrow \bm s^{(t)}_{k,m} - \bm f^{(t)}_{k,m}$;
    \ENDFOR
    \STATE Find stepsize $\rho^{(t)} \in (0,1]$ via Armijo backtracking to satisfy \eqref{eq:armijo_condition};
    \FOR{each antenna element $(k,m)$}
        \STATE Update: $\bm f^{(t+1)}_{k,m} \leftarrow \frac{\bm f^{(t)}_{k,m} + \rho^{(t)} \bm d^{(t)}_{k,m}}{\|\bm f^{(t)}_{k,m} + \rho^{(t)} \bm d^{(t)}_{k,m}\|_2}$;
    \ENDFOR
    \STATE Compute $R^{(t+1)} \leftarrow R(\{\bm f^{(t+1)}_{k,m}\})$;
    \STATE $t \leftarrow t + 1$.
\ENDWHILE
\end{algorithmic}
\end{algorithm}

\subsection{Overall AO-based Algorithm and Complexity Analysis}

The complete AO framework for joint beamforming and antenna–orientation optimization is summarized in Algorithm~\ref{alg:ao_framework}. Each outer iteration alternates between (i) updating the beamformers $\{\bm w_k\}$ with fixed orientations and (ii) updating the orientations with fixed beamformers. By construction, both substeps yield nondecreasing objective values. Consequently, the weighted sum-rate sequence $\{R^{(t)}\}$ is monotone nondecreasing. Since the feasible set is compact and the weighted sum-rate is bounded under the finite power budget, the sequence $\{R^{(t)}\}$ is bounded and therefore convergent.

{\textit{Complexity Analysis:}} For Algorithm~\ref{alg:ao_framework}, let $T_{\text{W}}$ be the inner WMMSE iterations, $T_{\text{FW}}$ the Frank--Wolfe steps, and $T_{\text{AO}}$ the outer AO iterations. In the beamforming block, one WMMSE sweep is dominated by forming per-user Gram matrices and solving $K$ linear systems of size $M$, yielding $\mathcal{O}(K^2 M^2 + K M^3)$ per sweep (the bisection on $\mu_k$ only affects constants), so the block costs $\mathcal{O}\!\big(T_{\text{W}}(K^2 M^2 + K M^3)\big)$ per outer iteration. In the orientation block, building all gradients and running the closed-form Frank--Wolfe oracle incurs $\mathcal{O}(K^2 M^2)$ per step, while Armijo backtracking adds lower-order evaluations, giving $\mathcal{O}(T_{\text{FW}} K^2 M^2)$ per outer iteration. Hence the overall complexity of Algorithm~\ref{alg:ao_framework} is approximately given by $\mathcal{O}\!\big(T_{\text{AO}}\big[T_{\text{W}}(K^2 M^2 + K M^3) + T_{\text{FW}} K^2 M^2\big]\big)$.

\begin{algorithm}[t] \small
\caption{The Overall AO-based Algorithm}
\label{alg:ao_framework}
\begin{algorithmic}[1]
\REQUIRE Feasible orientations $\{\bm f^{(0)}_{k,m}\}$, beamforming $\{\bm w_k^{(0)}\}$, tolerance $\epsilon > 0$, maximum iterations $T_{\max}$
\ENSURE Optimized beamforming vectors and antenna orientations
\STATE $t \leftarrow 0$;
\STATE Compute $R^{(0)} \leftarrow R(\{\bm w_k^{(0)}\}, \{\bm f^{(0)}_{k,m}\})$;
\WHILE{$t < T_{\max}$ and $\frac{|R^{(t)} - R^{(t-1)}|}{|R^{(t-1)}|} > \epsilon$}
    \STATE Fix $\{\bm f^{(t)}_{k,m}\}$, update $\{\bm w_k^{(t+1)}\}$ via WMMSE method;
    \STATE Fix $\{\bm w_k^{(t+1)}\}$, update $\{\bm f^{(t+1)}_{k,m}\}$ via Algorithm~\ref{alg:fw_orientation};
    \STATE Compute $R^{(t+1)} \leftarrow R(\{\bm w_k^{(t+1)}\}, \{\bm f^{(t+1)}_{k,m}\})$
    \STATE $t \leftarrow t + 1$,
\ENDWHILE
\end{algorithmic}
\end{algorithm}

\section{MRT/ZF Beamforming-Based Solution} 

The WMMSE beamforming block within the AO framework in the previous section delivers high performance but incurs substantial inner-loop complexity, especially with large arrays and many users. In contrast, classical beamforming schemes such as MRT and ZF provide closed-form expressions that are more tractable and computationally efficient. 

In this section, we fix the beamforming strategy to either MRT or ZF. As detailed below, substituting these explicit beamforming solutions into the original joint design removes the beamforming variables and reduces the task to a single-block orientation optimization. The resulting problems can be directly solved with a Frank–Wolfe–based algorithm, thus reducing complexity.

\textit{1) MRT Beamforming: }
The MRT scheme aligns each user’s beamforming vector with its instantaneous intended channel to maximize the received signal power, which is given by
\begin{align}
    \bm{w}_{\text{MRT},k}(\bm{F}_{k}) = \sqrt{P_{k,\text{max}}} \frac{\bm{h}_{k,k}(\bm{F}_{k})}{\| \bm{h}_{k,k}(\bm{F}_{k}) \|_2},\,\,\forall k.
\end{align}
With MRT beamforming, the achievable rate for each user $k$ is expressed as
\begin{align}\label{eq:SINR12}
    &R_{\text{MRT},k}(\{\bm{F}_{k}\}) \notag
    \\
    &= \log_{2}\Big(1+\frac{P_{k,\text{max}}\big\|\bm{h}_{k,k}\big\|_2^2}
    {\sum\limits_{n\neq k}^{K}P_{n,\text{max}} \big|\bm{h}_{n,k}^{H}\bm{h}_{n,n}\big|^2/\big\|\bm{h}_{n,n}\big\|_2^2 + \delta_k^2}\Big).
\end{align}
It follows that each user’s achievable rate depends solely on the antenna orientations.

\textit{2) ZF Beamforming:}
The ZF scheme projects each beamforming vector onto the null space of the unintended channels to suppress co-channel interference. For each transmitter $k$, stack the cross-link channels as
\begin{align}
    \bm H_{-k}(\bm F_k)
    \triangleq
    \big[\bm h_{k,1},\ldots,\bm h_{k,k-1},\bm h_{k,k+1},\ldots,\bm h_{k,K}\big].
\end{align}
Then, with $M\geq K$, the orthogonal projector onto their null space can be constructed as
\begin{align}
    \bm{P}^{\perp}_k(\bm F_k) = \bm I - \bm H_{-k}(\bm H_{-k}^H\bm H_{-k})^{-1}\,\bm H_{-k}^H,\,\,\forall k.
\end{align}
The ZF beamforming vectors are obtained as
\begin{align}
    \bm{w}_{\text{ZF},k}(\bm F_k) &= \sqrt{P_{k,\text{max}}} \frac{\bm{P}^{\perp}_k\bm{h}_{k,k}}{\| \bm{P}^{\perp}_k\bm{h}_{k,k} \|_2},\,\,\forall k.
\end{align}
With this choice, the achievable rate of each user $k$ is expressed as
\begin{align}
    R_{\text{ZF},k}(\bm{F}_{k})
    = \log_{2}\!\left(
      1+\frac{P_{k,\max} \| \bm{P}^{\perp}_k\bm{h}_{k,k} \|_2^2  }{\delta_k^{2}} 
    \right),
\end{align}
which shows that each user’s achievable rate depends solely on the antenna orientations.

\textit{3) Antenna Orientation Optimization:}
Under the MRT/ZF beamforming schemes, the original joint optimization Problem~\eqref{p1} reduces to
\begin{align}
    \max_{\{\bm{F}_k\}_{k=1}^K}
    \quad & \sum_{k=1}^{K} \alpha_k R_{\text{MRT/ZF},k} \label{pdfe1}\\
    \text{s.t.}\qquad 
    & \cos(\theta_{\max}) \le \bm{f}_{k,m}^{\!T}\bm{e}_z \le 1, \quad \forall k,m, \label{eq:orient_paramjhyygiu}\\
    & \|\bm{f}_{k,m}\|_2 = 1,\quad \forall k,m. \label{eq:unitnormdfgs}
\end{align}
The gradient of $\sum_{k=1}^{K} \alpha_k R_{\text{MRT/ZF},k}$ w.r.t. the antenna orientations $\{\bm{f}_{k,m}\}$ can be obtained via the chain rule (similar to Appendix~\ref{apdix1}) or by numerical estimation. Therefore, a Frank–Wolfe–based algorithm can be developed to tackle this problem. Since the approach closely mirrors Algorithm~1, we omit the detailed description here for brevity.

{\textit{Complexity Analysis:}} For the MRT-based beamforming scheme, the computational cost of each Frank–Wolfe step is dominated by the oracle evaluations, resulting in a per-step complexity of $\mathcal{O}(K^{2}M^{2})$ and an overall complexity of $\mathcal{O}(T_{\text{FW}}K^{2}M^{2})$. For the ZF-based beamforming scheme, an additional cost of $\mathcal{O}(K^{3}M + K^{4})$ per step is required to construct the projection matrices, leading to a total complexity of $\mathcal{O}\big(T_{\text{FW}}(K^{2}M^{2} + K^{3}M + K^{4})\big)$.

\section{Joint Optimization with Discrete Antenna Orientations}

\begin{figure*}[!t]
  \centering
  \subfloat[Uniform grid sampling]{%
    \includegraphics[width=0.45\linewidth]{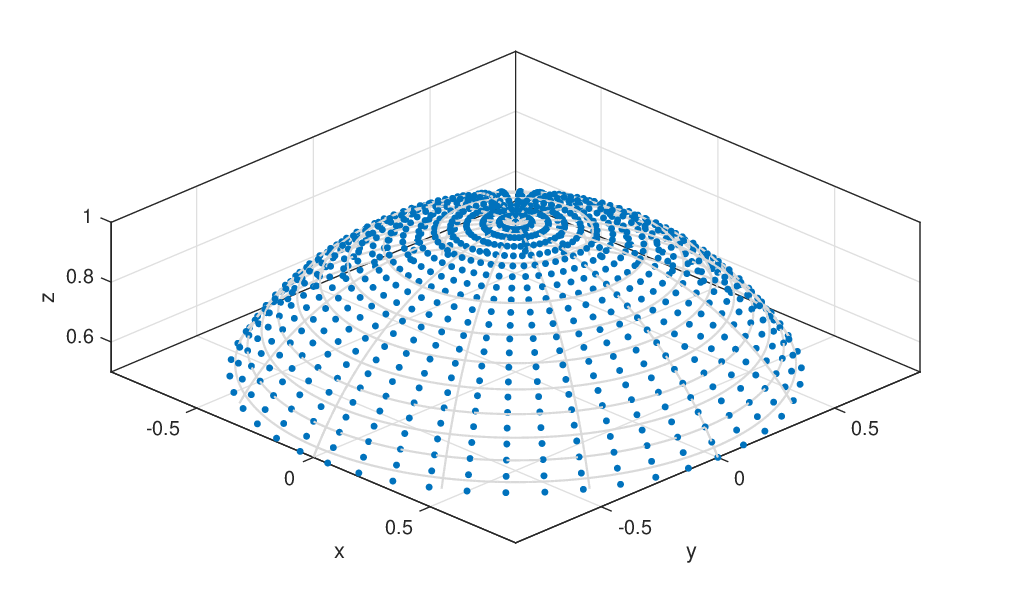}%
    \label{fig:uniform111}}
  \hfill
  \subfloat[Spherical Fibonacci sampling]{%
    \includegraphics[width=0.45\linewidth]{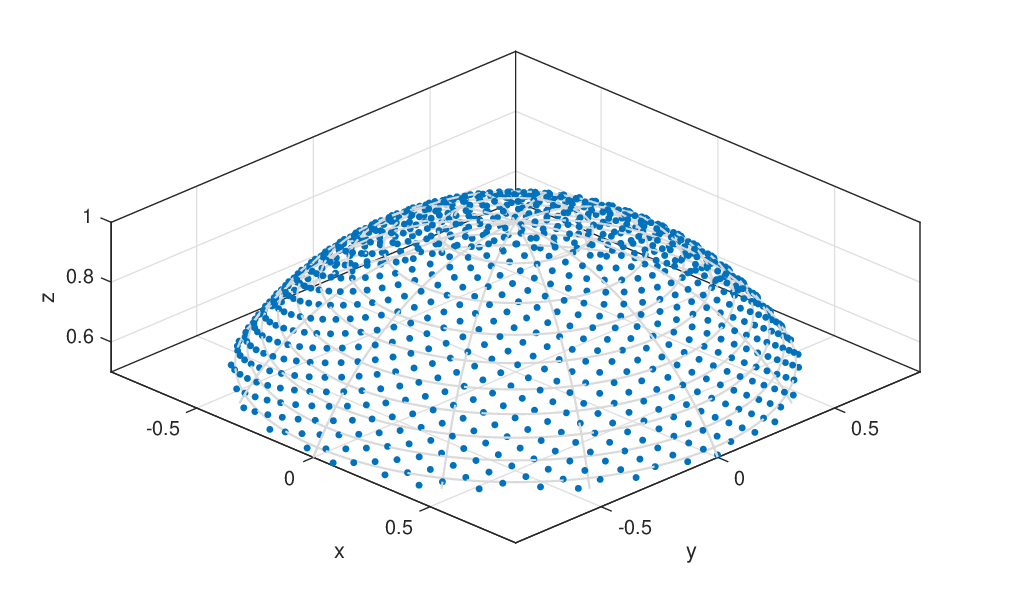}%
    \label{fig:fibo111}}
  \caption{Sampling schemes on a spherical cap (\(\theta_{\max}=\pi/3\)) for discrete orientations. Each circle indicates a reference boresight direction.}
  \label{fig:sampling_comp}
\end{figure*}

In practice, antenna orientations may not be adjusted with arbitrary precision owing to hardware constraints, such as finite-resolution mechanical rotation or codebook-based electronic control. This motivates the consideration of discrete orientation optimization, where each element can steer its boresight only toward a finite set of candidate directions. Compared with the continuous-orientation case, the resulting design problem is a more challenging mixed discrete–continuous optimization, in which beamforming must be jointly optimized with orientation selection. 

The feasible orientation domain of each antenna element is restricted to a spherical cap, i.e., \eqref{eq:single_cap}. An intuitive discretization strategy is to quantize the zenith and azimuth angles into uniform grids, e.g., 
\begin{align}
    &\theta \in \Big\{0,\ \tfrac{\theta_{\max}}{N_\theta-1},\ \dots,\ \theta_{\max}\Big\}, \\
    &\phi \in \Big\{-\pi,\ -\pi+\tfrac{2\pi}{N_\phi},\ \dots,\ \pi-\tfrac{2\pi}{N_\phi}\Big\},
\end{align}
where $N_\theta$ and $N_\phi$ denote the number of quantization levels for the zenith and azimuth angles, respectively, as illustrated in Fig.~\ref{fig:sampling_comp}(a). Although straightforward, this approach is not equal-area: the surface element on the sphere is $dA=\sin\theta\,d\theta\,d\phi$, so grid points become sparse near the cap rim and cluster around the pole, leading to non-uniform coverage. 

To overcome this limitation, we adopt spherical Fibonacci sampling, which generates nearly uniform and equal-area reference directions over the cap. 
For a codebook size $N_{\text{dir}}=N_\theta N_\phi$, 
the spherical Fibonacci samples are generated as \cite{r3}
\begin{align}
    \theta_i &= \arccos\left(1 - \frac{i+\tfrac12}{N_{\text{dir}}}\,\big(1-\cos\theta_{\max}\big)\right), \quad \forall i,\\
    \phi_i   &= 2\pi\,\text{frac}\!\left(\tfrac{i}{\varphi^2}\right),
    \quad \forall i,
\end{align}
where $\varphi=(1+\sqrt{5})/2$ is the golden ratio, and 
$\text{frac}(x)=x-\lfloor x \rfloor$ denotes the fractional part. Accordingly, the orientation codebook is then defined as
\begin{align}
    \mathcal{F} 
    = \big\{\, \bm f(\theta_i,\phi_i)\ :\ i=0,1,\dots,N_{\text{dir}}-1 \,\big\},
\end{align}
where $\bm f(\theta,\phi)$ is the unit-norm mapping in Eq.~\eqref{ffunc}.  
As illustrated in Fig.~\ref{fig:sampling_comp}(b), this Fibonacci construction ensures that each point represents an almost equal solid angle, avoids pole clustering, making it more suitable for both search and integration over the spherical cap. 

When each antenna element is restricted to a discrete set of orientations, the original problem in \eqref{p1} is reformulated as
\begin{align}
    \max_{\{\bm{w}_k,\bm{F}_k\}_{k=1}^K}
    \quad & \sum_{k=1}^{K} \alpha_k R_k \label{p11}\\
    \text{s.t.}\qquad \,\,
    & \|\bm{w}_k\|_2^2 \le P_{k,\max}, \quad \forall k, \label{eq:pk1}\\
    & \bm f_{k,m} \in \mathcal{F}, \quad \forall k,m.  \label{eq:discreconst}
\end{align}
This problem is challenging due to the discrete orientation constraint, which renders the formulation nonconvex. Moreover, exhaustive search over $|\mathcal{F}|^{KM}$ possibilities scales exponentially with the array size and is thus computationally infeasible, necessitating low-complexity approximate algorithms.

\begin{algorithm}[t] \small
\caption{CEM-based Algorithm}
\label{alg:cem_framework}
    \begin{algorithmic}[1]
        \REQUIRE Sample size $S$, elite fraction $\rho$, smoothing $\tau$, Maximum iterations $T_{\max}$, codebook $\mathcal F$, initial $\{\bm w_k^{(0)}\}$
        \ENSURE Optimized orientations $\bm F^{\text{best}}$ and beamforming $\{\bm w_k^{\text{best}}\}$
        \STATE Let $J$ denote the number of orientation variables (e.g., $J=KM$). Initialize categorical pmfs $\{p_j^{(0)}\}_{j=1}^{J}$ over $\mathcal F$ (e.g., uniform), denoted as $\{\text{Cat}(p_j^{(0)})\}_{j=1}^J$.
        \STATE \textbf{Initialization:} Sample $\bm F^{(0)} \sim \prod_{j=1}^J \text{Cat}(p_j^{(0)})$. Given $\bm F^{(0)}$, solve for $\{\bm w_k^{(0)}\}$ by WMMSE; set $R^{(0)} \!\leftarrow\! R(\{\bm w_k^{(0)}\}, \bm F^{(0)})$ and $(\bm F^{\text{best}},\{\bm w_k^{\text{best}}\},R^{\text{best}})\!\leftarrow\!(\bm F^{(0)},\{\bm w_k^{(0)}\},R^{(0)})$.
        \FOR{$t=1$ to $T_{\max}$}
          \STATE \textbf{Sampling:} Draw $\{\bm F^{(s)}\}_{s=1}^{S} \sim \prod_{j=1}^J \text{Cat}\!\big(p_j^{(t-1)}\big)$. For each $s$, solve for $\{\bm w_k^{(s)}\}$ by WMMSE; then compute $R^{(s)} \leftarrow R(\{\bm w_k^{(s)}\}, \bm F^{(s)})$.
          \STATE \textbf{Elite selection:} Let $\mathcal{E}$ contain the indices of the top $\lceil \rho S \rceil$ candidates ranked by $R^{(s)}$. Then, for each variable $j$ and each $f \in \mathcal{F}$, compute the empirical frequency: $q_j(f)=\frac{1}{|\mathcal E|}\sum_{s\in\mathcal E}\bm 1\{F^{(s)}_j=f\}$.
          \STATE \textbf{Update:} $p_j^{(t)} \leftarrow (1-\tau)\,p_j^{(t-1)}+\tau\,q_j$, for $j=1,\dots,J$. If $\max_s R^{(s)} > R^{\text{best}}$ with maximizer $s^\star$, set $(\bm F^{\text{best}},\{\bm w_k^{\text{best}}\},R^{\text{best}})\leftarrow(\bm F^{(s^\star)},\{\bm w_k^{(s^\star)}\},R^{(s^\star)})$.
        \ENDFOR
        \STATE \textbf{Output:} $\bm F^{\text{best}}$ and $\{\bm w_k^{\text{best}}\}$.
    \end{algorithmic}
\end{algorithm}

To tackle this issue, we employ the cross-entropy method (CEM), a model-based stochastic search well suited to large discrete spaces \cite{r4}. CEM iteratively learns a probabilistic model that concentrates its probability mass on high-quality regions of the feasible set. Concretely, we maintain for each orientation variable a categorical distribution over the codebook $\mathcal{F}$. Each CEM iteration proceeds in three steps:
(i) \textit{Sampling}: draw $S$ orientation sets from the current categorical distributions and, for each sample, solve the WMMSE subproblem to obtain $\{\bm w_k\}$ and its weighted sum-rate;
(ii) \textit{Elite selection}: select the top $\lceil \rho S\rceil$ samples ranked by the weighted sum-rate;
(iii) \textit{Update}: move each categorical probability mass function (pmf) toward the elites via an exponential moving average. Because sampling is performed directly on $\mathcal{F}$, the discrete constraint \eqref{eq:discreconst} holds by construction. The complete algorithmic procedure is summarized in Algorithm 3. In practice, CEM explores the discrete space efficiently and typically achieves high-quality approximations to \eqref{p11}.

\textbf{\textit{Complexity Analysis:}} For Algorithm~\ref{alg:cem_framework}, let \(S\) be the batch size, \(T_{\text{W}}\) the WMMSE inner iterations, \(T_{\max}\) the total CEM rounds, and \(|\mathcal F|\) the codebook size. In each CEM iteration, the dominant cost is solving WMMSE \(S\) times, each of complexity \(\mathcal{O}\!\big(T_{\text{W}}(K^{2}M^{2}+KM^{3})\big)\); the remaining overheads are sampling \(\mathcal{O}(SJ)\), elite selection \(\mathcal{O}(S\log S)\), and updating the categorical distributions \(\mathcal{O}(J|\mathcal F|)\). Hence the overall computational complexity is given by \(\mathcal{O}\!\big(T_{\max} S T_{\text{W}}(K^{2}M^{2}+KM^{3})\big)\).

\section{Numerical Results}
In this section, we evaluate the performance of the proposed algorithms and schemes for MISO interference channels. Unless otherwise specified, the simulation parameters are summarized in Table~\ref{tab:sim_params}. The 3D topology is randomly generated per trial (units in meters) as follows. Transmitter array centers are placed at $\bm t_k=[x,\,20,\,0]$ with $x\!\sim\!\mathcal U[0,80]$. Scatterer clusters are located at $\bm s_q=[x,\,6,\,z]$ with $x\!\sim\!\mathcal U[0,100]$ and $z\!\sim\!\mathcal U[20,40]$. Receiver positions are $\bm u_k=[x,\,2,\,z]$ with $x\!\sim\!\mathcal U[0,100]$ and $z\!\sim\!\mathcal U[80,100]$. Each cluster is assigned a random phase $\chi_q\!\sim\!\mathcal U[0,2\pi)$ and an average scattering power $\eta_q\!=\!0.5$. Unless stated otherwise, all antenna-element orientations are initialized to point along the $+z$ axis.  We assume identical per-UPA power budgets, i.e., $P_{k,\max}=P_{\max},\forall k$.

\begin{table}[t] 
\centering
\caption{Simulation parameters}
\label{tab:sim_params}
\renewcommand{\arraystretch}{1.08}
\begin{tabular}{c || c}
\hline\hline
\textbf{Parameter} & \textbf{Value} \\
\hline
    Number of Tx--Rx pairs, $K$ & $4$ \\
    Number of scatterer clusters, $Q$ & $6$ \\
    Wavelength, $\lambda$ & $0.125~\text{m}$ \\
    Inter-element spacing, $d$ & $2\lambda$ \\
    The number of elements per UPA, $M $  & $4$ \\
    Noise power at user, $\delta_k^2,\forall k$ & $-80~\text{dBm}$ \\
    Maximum zenith angle, $\theta_{\max}$ & $\pi/3$ \\
    User weights, $\alpha_k,\forall k$ & $1$ \\
\hline\hline
\end{tabular}
\end{table}

\subsection{Convergence Behavior}

\begin{figure}[t]
  \centering
  \subfloat[AO-based Algorithm 2 for Continuous Orientation]{
    \includegraphics[width=0.9\linewidth]{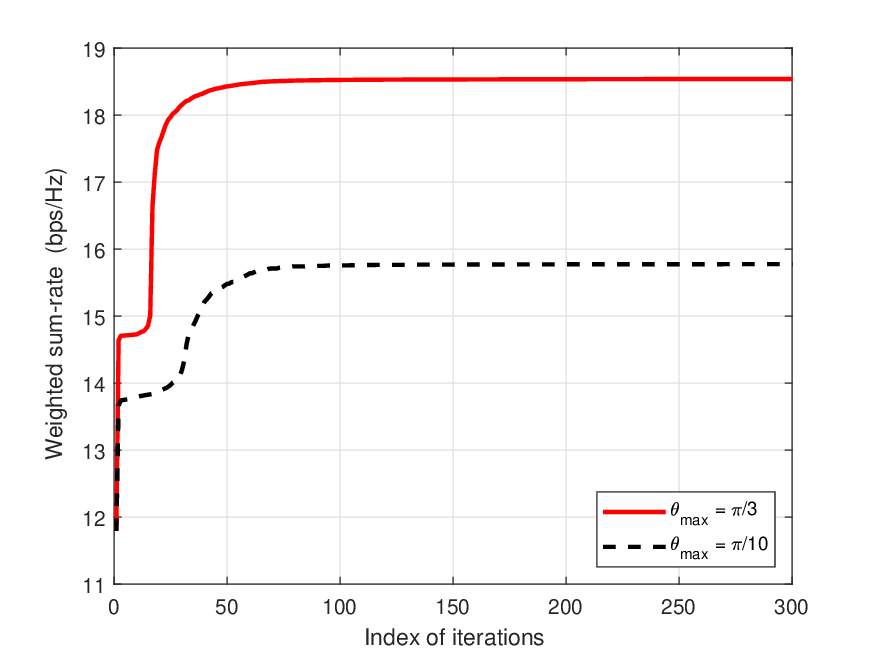}
    \label{fig:uniform}
  } \\
  \subfloat[CEM-based Algorithm 3 for Discrete Orientation]{
    \includegraphics[width=0.9\linewidth]{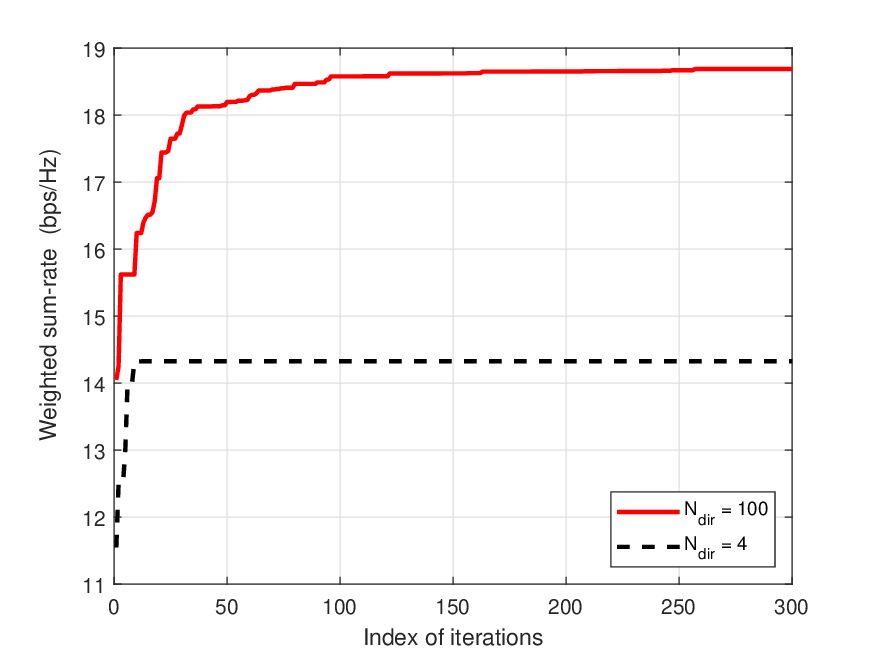}
    \label{fig:fibo}
  }
  \caption{Convergence behavior of the proposed algorithms.}
  \label{fig:conver}
\end{figure}

Before the performance comparison, Fig.~\ref{fig:conver} depicts the convergence behavior of the proposed algorithms. In Fig.~\ref{fig:conver}(a), the AO-based algorithm for continuous orientations is evaluated under maximum zenith angle $\theta_{\max}\in\{\pi/3,\,\,\pi/10\}$. In all cases, the weighted sum-rate increases monotonically and converges within several tens of iterations. A larger $\theta_{\max}$ leads to a higher final weighted sum-rate, as the expanded feasible region facilitates improved beam alignment and interference management. In Fig.~\ref{fig:conver}(b), the CEM-based algorithm for discrete orientations (with $\theta_{\max}=\pi/3$) is tested with codebook sizes $N_{\text{dir}}\in\{4,\,100\}$. The best achievable objective value shows steady improvement and eventually converges across all quantization levels. Finer quantization consistently yields superior final performance due to reduced codebook mismatch w.r.t. the continuous manifold.

\subsection{Performance Comparison Under Different Beamforming Schemes}
In this subsection, we compare the weighted sum-rate performance achieved by various beamforming schemes combined with rotation optimization methods, namely the algorithms proposed in Sections III and IV. Specifically, the following schemes are considered:

\begin{itemize}
  \item \textbf{WMMSE beamforming + RA:} 
  The AO-based solution proposed in Section III, where the beamforming vectors are optimized via the WMMSE method, and the antenna orientation is updated by a Frank–Wolfe-based approach.

  \item \textbf{MRT beamforming + RA:}
  The solution presented in Section IV, which applies the MRT structure for beamforming and optimizes the antenna orientation via the Frank–Wolfe-based method.

  \item \textbf{ZF beamforming + RA:}
  The solution presented in Section IV, which applies the ZF structure for beamforming and optimizes the antenna orientation via the Frank–Wolfe-based method.

  \item \textbf{Baseline: WMMSE/MRT/ZF beamforming:}
  The antenna orientation is fixed to the $+z$ direction without rotation, while the beamforming vectors are optimized using the WMMSE, MRT, or ZF method, respectively.
\end{itemize}

\begin{figure}[t]
	\begin{center}
		\includegraphics[width=0.45\textwidth]{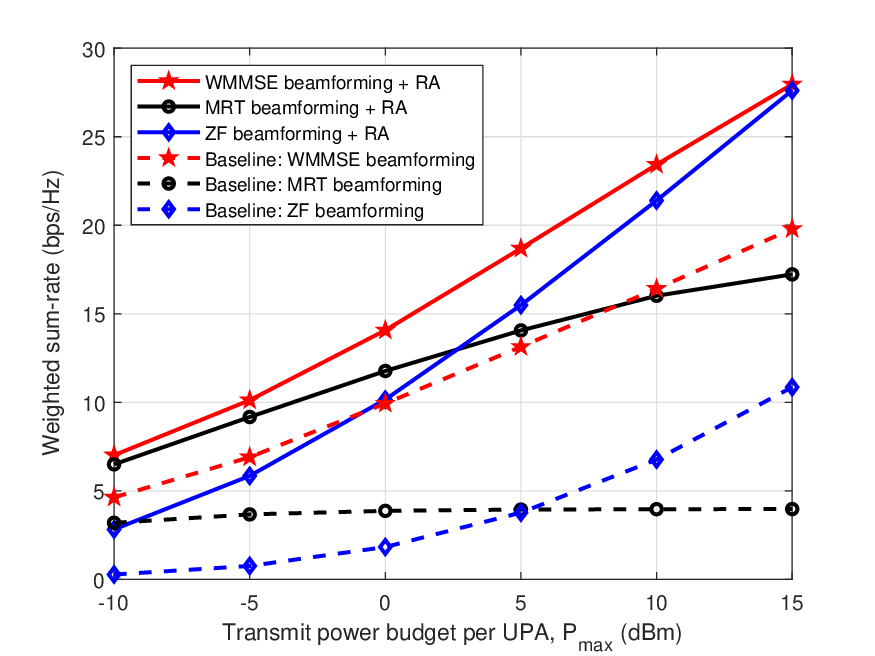}
		\caption{Weighted sum-rate vs. transmit power budget per UPA.}
        \label{fig:simulVesusPth}
	\end{center}
\end{figure}

Fig.~\ref{fig:simulVesusPth} illustrates the weighted sum-rate versus the per-UPA power budget \(P_{\max}\) with antenna directivity factor $p=4$. All methods improve as power increases, and the three RA-enabled schemes consistently and significantly outperform their fixed-orientation counterparts across the entire range, demonstrating the clear advantage of adaptive boresight control. Among RA-enabled schemes, \emph{WMMSE+RA} achieves the highest throughput at all power levels. In the low transmit-power regime (noise-limited), \emph{MRT+RA} performs nearly on par with \emph{WMMSE+RA}, as MRT already captures most of the array gain while RA steering aligns the main lobes toward intended users, leaving limited headroom for additional interference management. As \(P_{\max}\) increases and the operation becomes more interference-limited, \emph{ZF+RA} improves most rapidly owing to its interference-suppression capability, narrowing the gap to \emph{MRT+RA} and potentially surpassing it in strongly interference-dominated settings. Throughout, \emph{WMMSE+RA} remains the overall best performer. The widening separation between RA-enabled and fixed-orientation curves with increasing \(P_{\max}\) further underscores that orientation optimization is particularly beneficial in the interference-limited regime, where it simultaneously strengthens desired links and mitigates inter-user interference.

\begin{figure}[t]
	\begin{center}
		\includegraphics[width=0.45\textwidth]{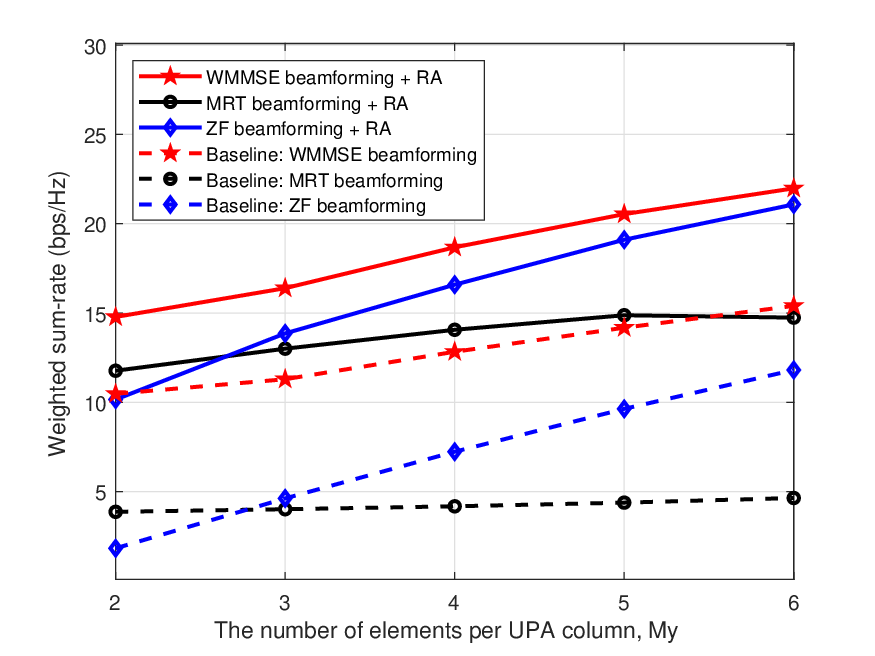}
		\caption{Weighted sum-rate vs. the number of elements per UPA column.}
        \label{fig:simulVersusMy}
	\end{center}
\end{figure}

Fig.~\ref{fig:simulVersusMy} depicts the weighted sum-rate versus the number of elements per UPA column \(M_y\) at \(P_{\text{max}}=0~\mathrm{dBm}\), with \(M_x=2\) and \(M_y\in\{2,3,4,5,6\}\) yielding \(M\in\{4,6,8,10,12\}\) elements per UPA. All methods benefit from larger arrays, and the RA-enabled designs consistently dominate their fixed-orientation counterparts for every \(M_y\). Within the RA-enabled family, \emph{WMMSE+RA} remains the top performer across all sizes. \emph{MRT+RA} is closest at small arrays (noise-limited regime) but its gain saturates as \(M\) grows and interference becomes more prominent. In contrast, \emph{ZF+RA} exhibits the steepest growth with \(M_y\) because the additional spatial degrees of freedom improve channel conditioning and enable sharper interference suppression, narrowing its gap to \emph{WMMSE+RA}. Overall, increasing \(M_y\) enlarges the aperture and directivity, allowing orientation optimization to more effectively steer main lobes toward intended users and reduce cross-link coupling, and the performance gap between RA-enabled and non-RA baseline schemes widens with array size.

\subsection{Performance Comparison Under Different RA Configurations}
In this subsection, we evaluate the weighted sum-rate performance under different RA configurations, incorporating parameters such as the antenna directivity factor, maximum zenith angle, and orientation codebook size. Meanwhile, the following schemes are compared:
\begin{itemize}
  \item \textbf{Cont.-RA, AO solution:}
  The orientation of each antenna can be adjusted continuously, as described in Section~III. The antenna orientations and beamforming vectors are jointly optimized via the AO framework (Algorithm~\ref{alg:ao_framework}).

  \item \textbf{Disc.-RA, CEM solution:}
  The orientation of each antenna is constrained to a discrete set, as described in Section~V. The joint beamforming and orientation selection is optimized using the CEM-based algorithm (Algorithm~\ref{alg:cem_framework}).

  \item \textbf{Disc.-RA, Proj. solution:}
  The orientation of each antenna is restricted to a discrete set, as described in Section~V. The orientations and beamformers are first optimized via AO method (Algorithm~\ref{alg:ao_framework}), after which the resulting orientations are projected to the nearest available discrete points.

  \item \textbf{Baseline: fixed orientations:}
  All antenna boresights are fixed to the $+z$ axis without rotation, and only the beamformers are optimized via WMMSE method.

  \item \textbf{Baseline: isotropic elements:}
  Each element is isotropic (i.e., $p=0$), and the beamformers are optimized via WMMSE method.
\end{itemize}

\begin{figure}[t]
	\begin{center}
		\includegraphics[width=0.45\textwidth]{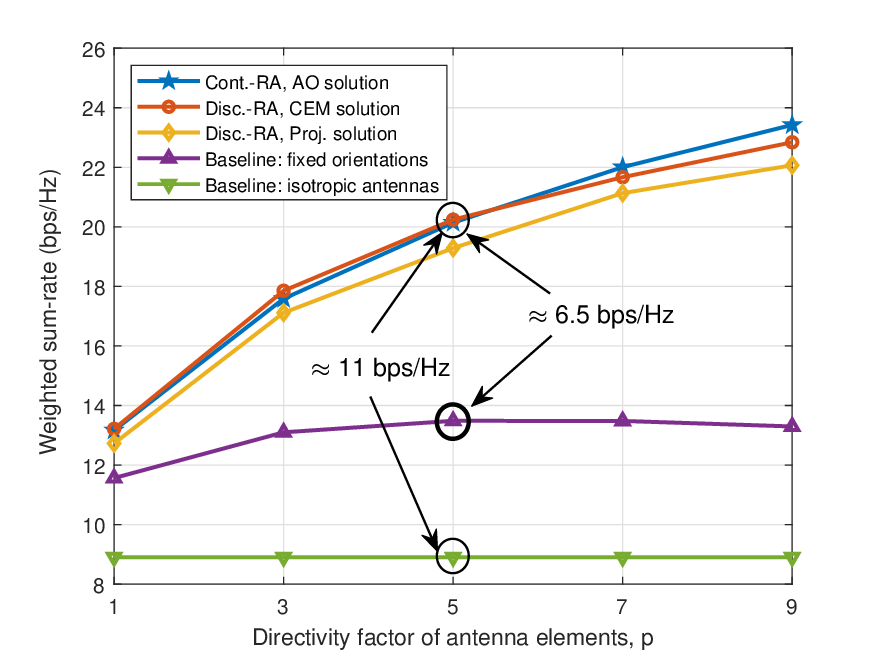}
		\caption{Weighted sum-rate vs. directivity factor of the antenna.}
            \label{fig:simulVersusP}
	\end{center}
\end{figure}

\begin{figure}[t]
	\begin{center}
		\includegraphics[width=0.45\textwidth]{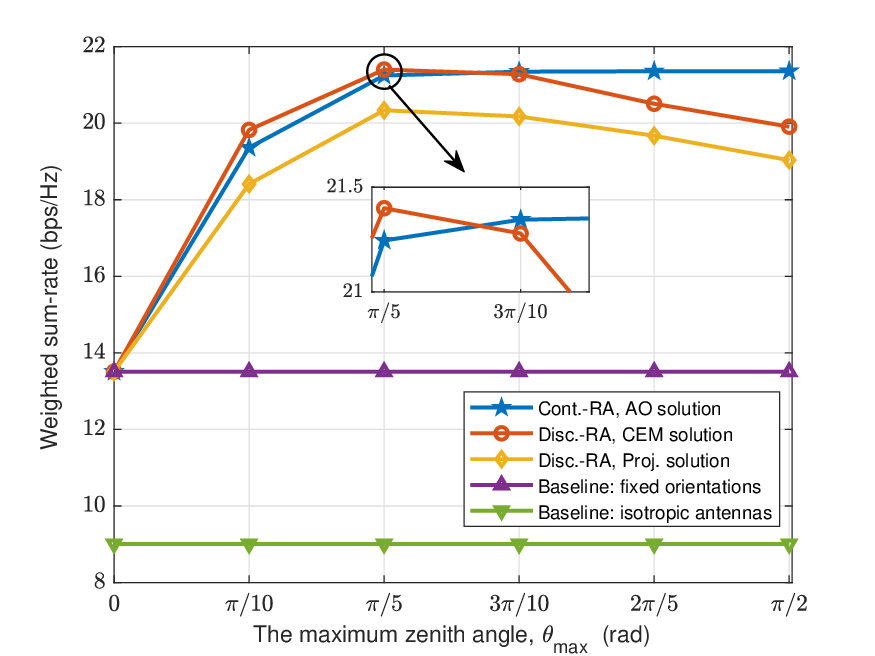}
		\caption{Weighted sum-rate vs. the maximum zenith angle.}
        \label{fig:simulVersusTheta}
	\end{center}
\end{figure}

\begin{figure}[t]
	\begin{center}
		\includegraphics[width=0.45\textwidth]{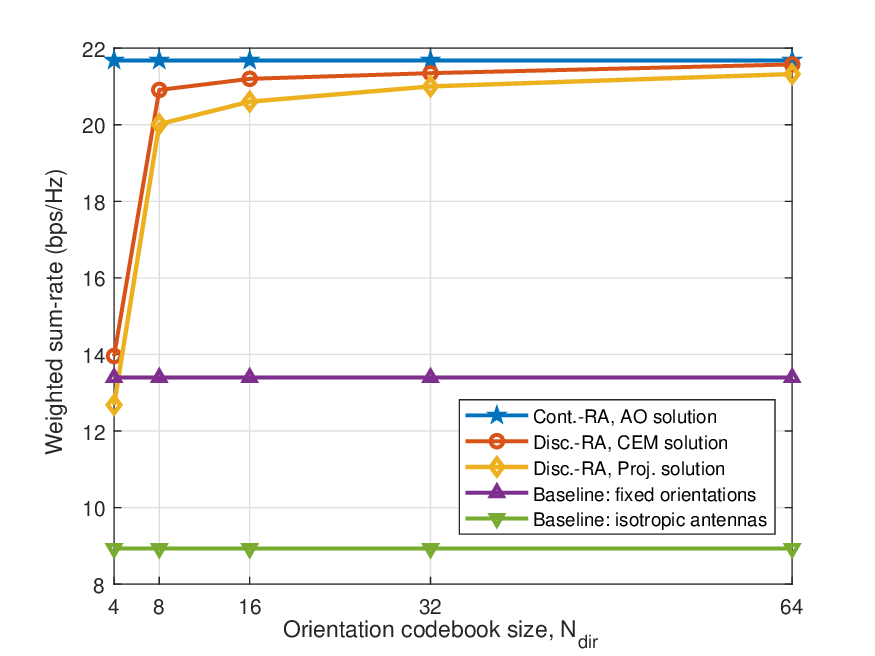}
		\caption{Weighted sum-rate vs. quantization codebook size.}
        \label{fig:simulVesusNdir}
	\end{center}
\end{figure}

Fig.~\ref{fig:simulVersusP} plots the weighted sum-rate versus the antenna directivity factor at \(P_{\text{max}}=5\,\mathrm{dBm}\). For the discrete RA designs we adopt a spherical-cap codebook with size \(N_{\text{dir}}=25\). As \(p\) increases, all RA-enabled schemes improve markedly, and the gap widens because higher element directivity strengthens the benefit of steering main lobes toward intended users while deflecting cross-link leakage. For example, at \(p=5\) the continuous-RA AO solution provides gains of about \(6.5\)~bps/Hz over the fixed-orientation baseline and \(11\)~bps/Hz over the isotropic-antenna baseline. For the discrete-orientation case, the CEM-based design consistently outperforms the nearest-projection rule, since stochastic exploration with elite averaging searches the nonconvex orientation landscape more effectively than a single projection. Notably, at small \(p\) the discrete CEM solution can even slightly exceed the continuous AO solution, because broad beams make the objective surface flatter, and the sampling-based CEM can jump across attraction basins where the gradient-driven AO method may settle into a conservative local optimum.

Fig.~\ref{fig:simulVersusTheta} plots the weighted sum-rate versus the maximum zenith angle with $p=6$ and $P_{\text{max}} = 5~\text{dBm}$. When \(\theta_{\max}=0\), the RA curves coincide with the fixed-orientation baseline. Allowing a small tilt up to \(\pi/10\) yields a sharp improvement for all RA-enabled designs, and the gains largely saturate around \(\theta_{\max}\!\approx\!\pi/5\). Beyond this point, the continuous-RA AO solution remains near its peak, whereas the two discrete solutions show a mild decline as the spherical cap widens while the codebook size is fixed \((N_{\text{dir}}=25)\), which lowers the angular sampling density and complicates precise boresight locking. Across the entire range, the RA-enabled schemes stay well above both baselines, and the isotropic-antenna baseline is essentially flat. Between the discrete methods, the CEM-based design consistently exceeds the nearest-projection rule. At small \(\theta_{\max}\), the Discrete-RA CEM curve can even slightly surpass the Continuous-RA AO curve, because a tight cap often places the optimum near the boundary and the sampling-based search is more likely to hit such boundary-aligned orientations, whereas the gradient-based AO solution tends to be conservative within the feasible set.

Fig.~\ref{fig:simulVesusNdir} plots the weighted sum-rate versus the orientation codebook size. The continuous-RA AO curve is flat since it does not depend on a codebook. The discrete RA curves rise sharply from \(N_{\text{dir}}=4\) to \(8\) and then approach a plateau close to the continuous upper envelope. By \(N_{\text{dir}}\ge 16\), the CEM solution is within a fraction of a bps/Hz of the AO curve, while the nearest-projection solution is slightly lower yet already near saturation. With a very small codebook of four directions, the projection method can even fall below the fixed-orientation baseline because of severe quantization, whereas CEM still delivers a large gain by jointly exploring codeword combinations. The two baselines remain essentially flat as \(N_{\text{dir}}\) varies. These trends indicate that a modest codebook size from \(N_{\text{dir}}=8\) to \(16\) captures most of the RA benefit, and that the proposed CEM-based algorithm is more robust than the nearest-projection method when angular resolution is coarse.

\section{Conclusions}

This paper investigated the application of rotatable antennas (RAs) for spectrum sharing in multiple-input single-output (MISO) interference channels. A weighted sum-rate maximization problem was formulated under spherical-cap constraints, jointly optimizing transmit beamforming and per-antenna orientation. An alternating optimization (AO) framework was developed, integrating WMMSE-based beamforming with Frank–Wolfe-based orientation updates incorporating Armijo backtracking. To reduce computational complexity, MRT and ZF beamforming strategies were also discussed, simplifying the problem to a single-block orientation optimization solved via the Frank–Wolfe method. Furthermore, practical finite-resolution actuation was implemented using spherical Fibonacci codebooks and a cross-entropy-based method.

Simulation results confirm that the incorporation of rotatable antennas substantially improves the weighted sum-rate compared to conventional beamforming systems. The proposed AO approach achieves the highest performance, while the low-complexity alternatives, MRT+RA and ZF+RA, attain near-optimal performance under low and high transmit power regimes, respectively. Performance gains increase with both element directivity and array size, with even limited tilt flexibility providing significant improvements that saturate at moderate zenith angles. For discrete actuation, a spherical Fibonacci codebook combined with CEM optimization closely matches the performance of continuous orientation with small codebook sizes and consistently outperforms the simple nearest-projection method.

\begin{appendix}

\subsection{Derivation for
\texorpdfstring{$\nabla_{\bm f_{k,m}}(\sum_{\ell=1}^K \alpha_\ell R_\ell)$}
{gradient w.r.t. f\_\{k,m\}}}

\label{apdix1}

For notational convenience, define for each user $\ell$, $S_\ell \triangleq \big|\bm h_{\ell,\ell}^H\bm w_\ell\big|^2$, and $I_\ell \triangleq \sum_{n\neq \ell}\big|\bm h_{n,\ell}^H\bm w_n\big|^2+\delta_\ell^2$. By the chain rule,
\begin{align}
    \nabla_{\bm f_{k,m}} (\alpha_\ell R_\ell)
    = \frac{\alpha_\ell}{\ln 2}\,
      \frac{I_\ell\,\nabla_{\bm f_{k,m}} S_\ell - S_\ell\,\nabla_{\bm f_{k,m}} I_\ell}
           {I_\ell\,(I_\ell+S_\ell)}.
    \label{eq:gen-dR}
\end{align}
Only channels radiated from transmitter $k$ depend on $\bm f_{k,m}$. Hence,
\begin{align}
    \nabla_{\bm f_{k,m}} S_\ell &=
    \begin{cases}
    2\,\Re\!\Big((\bm{h}_{k,k}^{H}\bm{w}_{k})^{*}\,
       \nabla_{\bm f_{k,m}} (\bm{h}_{k,k}^{H}\bm{w}_{k})\Big), & \ell=k,\\[2pt]
    \bm 0, & \ell\neq k,
    \end{cases}
    \label{eq:dS}\\
    \nabla_{\bm f_{k,m}} I_\ell &=
    \begin{cases}
    \bm 0, & \ell=k,\\[2pt]
    2\,\Re\!\Big((\bm{h}_{k,\ell}^{H}\bm{w}_{k})^{*}\,
       \nabla_{\bm f_{k,m}} (\bm{h}_{k,\ell}^{H}\bm{w}_{k})\Big), & \ell\neq k.
    \end{cases}
    \label{eq:dI}
\end{align}
Substituting \eqref{eq:dS}–\eqref{eq:dI} into \eqref{eq:gen-dR}, we obtain
\begin{align}
    &\nabla_{\bm f_{k,m}} (\alpha_k R_k) 
    = \frac{2\alpha_k}{\ln 2}\,
       \frac{\Re\!\Big((\bm{h}_{k,k}^{H}\bm{w}_{k})^{*}\,
           \nabla_{\bm f_{k,m}} (\bm{h}_{k,k}^{H}\bm{w}_{k})\Big)}{I_k+S_k}, 
    \label{eq:dRk}\\
    &\nabla_{\bm f_{k,m}} (\alpha_\ell R_\ell) 
    = -\,\frac{2\alpha_\ell S_\ell}{\ln 2\, I_\ell\,(I_\ell+S_\ell)} \notag \\ 
       &\qquad \qquad  \times \Re\!\Big((\bm{h}_{k,\ell}^{H}\bm{w}_{k})^{*}\,
           \nabla_{\bm f_{k,m}} (\bm{h}_{k,\ell}^{H}\bm{w}_{k})\Big),
    \,\, \forall \ell\neq k.
    \label{eq:dRl}
\end{align}
Since $\bm h_{k,k}^{H}\bm w_k
=\sum_{i=1}^{M} h_{k,i,k}^{*}\,w_{k,i}$, and only the $m$-th entry $h_{k,m,k}$ depends on $\bm f_{k,m}$, it follows that, for $\ell =k$,
\begin{align}
    \nabla_{\bm f_{k,m}}\!\big(\bm h_{k,k}^{H}\bm w_k\big)
    &= w_{k,m}\,\big(\nabla_{\bm f_{k,m}} h_{k,m,k}\big)^{*}.
    \label{eq:grad_hHw_form1}
\end{align}
Similarly, for $\ell\neq k$,
\begin{align}
    \nabla_{\bm f_{k,m}}\!\big(\bm h_{k,\ell}^{H}\bm w_k\big)
    = w_{k,m}\,\big(\nabla_{\bm f_{k,m}} h_{k,m,\ell}\big)^{*}.
    \label{eq:grad_hHw_form2}
\end{align}
Let $\hat{\bm{d}}_{k,m,n}=\bm u_n-\bm t_{k,m}$, $r_{k,m,n}=\|\hat{\bm{d}}_{k,m,n}\|$, $\bm o_{k,m,q}=\bm s_q-\bm t_{k,m}$, $\tilde r_{k,m,q}=\|\bm o_{k,m,q}\|$, $\hat r_{q,n}=\|\bm s_q-\bm u_n\|$, and $C_0\triangleq \sqrt{\beta_0\kappa_{\max}}$. Then we have,
\begin{align}
    \nabla_{\bm f_{k,m}} h^{\text{LoS}}_{k,m,n}
    &= C_0\,\frac{p[\bm f_{k,m}^{T}\hat{\bm{d}}_{k,m,n}]_{+}^{p-1}}{r_{k,m,n}^{\,p+1}}\hat{\bm{d}}_{k,m,n}\;e^{-j\frac{2\pi}{\lambda}r_{k,m,n}},\\
    \nabla_{\bm f_{k,m}} h^{\text{NLoS}}_{k,m,n}
    &= \sum_{q=1}^{Q} \frac{C_0\sqrt{\sigma_q/4\pi}\,p}{\tilde r_{k,m,q}^{\,p+1}\,\hat r_{q,n}}\;
       [\bm f_{k,m}^{T}\bm o_{k,m,q}]_{+}^{p-1} \notag \\
       &\qquad \quad \times\bm o_{k,m,q}\;
       e^{-j\frac{2\pi}{\lambda}(\tilde r_{k,m,q}+\hat r_{q,n})+j\chi_q},\\
    \nabla_{\bm f_{k,m}} h_{k,m,n}
    &= \nabla_{\bm f_{k,m}} h^{\text{LoS}}_{k,m,n}
     + \nabla_{\bm f_{k,m}} h^{\text{NLoS}}_{k,m,n}.
    \label{eq:grad_h_total}
\end{align}
We adopt $[x]_+^{\,p-1}=0$ whenever $[x]_+=0$, i.e., zero Clarke subgradient at the visibility boundary. Substituting \eqref{eq:grad_hHw_form1}–\eqref{eq:grad_hHw_form2} and \eqref{eq:grad_h_total} into
\eqref{eq:dRk}–\eqref{eq:dRl}, and using $\Re\{a^{*}b^{*}\}=\Re\{ab\}$, we obtain

\vspace{-10pt}

{\small 
\begin{align}
    &\nabla_{\bm f_{k,m}}\!\Big(\sum_{\ell=1}^K \alpha_\ell R_\ell\Big)
    = \frac{2}{\ln 2}\,
    \Re\!\Bigg\{
    w_{k,m}^{*}\Bigg[
    \frac{\alpha_k\,\bm h_{k,k}^{H}\bm w_k}{I_k+S_k}\,
    \nabla_{\bm f_{k,m}} h_{k,m,k} \notag \\
    &\qquad \quad   - \sum_{\ell\neq k}\frac{\alpha_\ell S_\ell}{I_\ell\,(I_\ell+S_\ell)}\,
    \big(\bm h_{k,\ell}^{H}\bm w_k\big)\,\nabla_{\bm f_{k,m}} h_{k,m,\ell}
    \Bigg]\Bigg\}. 
    \label{eq:sumrate-grad-finald}
\end{align}
}

\subsection{The proof of Proposition 1}
\label{apdix20}

\begin{IEEEproof}
    Since $s^{(t)}_{k,m}$ is the optimal solution to oracle \eqref{eq:linear_oracle}, and the current point $\bm f^{(t)}_{k,m}\in\mathcal C_{\rm cap}$ is feasible, we have $\langle \bm T^{(t)}_{k,m}\bm g^{(t)}_{k,m},\,\bm s^{(t)}_{k,m}\rangle
    \ge \langle \bm T^{(t)}_{k,m}\bm g^{(t)}_{k,m},\,\bm f^{(t)}_{k,m}\rangle$. Further, because $\bm T^{(t)}_{k,m}=\bm I-\bm f^{(t)}_{k,m}(\bm f^{(t)}_{k,m})^{\!T}$ is symmetric and $\bm T^{(t)}_{k,m}\bm f^{(t)}_{k,m}=\bm 0$, we have
    $\langle \bm T^{(t)}_{k,m}\bm g^{(t)}_{k,m},\,\bm f^{(t)}_{k,m}\rangle
    =\langle \bm g^{(t)}_{k,m},\,\bm T^{(t)}_{k,m}\bm f^{(t)}_{k,m}\rangle=0$.
    Summing over all $(k,m)$ yields $\sigma^{(t)}=\sum_{k,m}\langle \bm T^{(t)}_{k,m}\bm g^{(t)}_{k,m},\,\bm s^{(t)}_{k,m}\rangle\ge 0$.
\end{IEEEproof}

\subsection{The proof of Theorem 1}
\label{apdix2}

\begin{IEEEproof}
With Eq. \eqref{eq:update_rule} and a stepsize $\rho$, the orientations are updated at iteration $t$ as

\vspace{-10pt}

{\small 
\begin{align}
\label{upeq}
\bm f^{(t+1)}_{k,m}(\rho)
&= \frac{\bm f^{(t)}_{k,m} + \rho \bm d^{(t)}_{k,m}}
       {\big\|\bm f^{(t)}_{k,m} + \rho \bm d^{(t)}_{k,m}\big\|_2} \notag \\
&= \frac{(1-\rho)\bm f^{(t)}_{k,m}+\rho\,\bm s^{(t)}_{k,m}}
       {\big\|(1-\rho)\bm f^{(t)}_{k,m}+\rho\,\bm s^{(t)}_{k,m}\big\|_2},\ \forall k,m.
\end{align}
}

\vspace{-5pt}

\noindent By construction, $\{\bm f^{(t+1)}_{k,m}(\rho)\}$ is feasible for all $\rho\in[0,1]$. Let $\Phi(\rho)=R(\{\bm f^{(t+1)}_{k,m}(\rho)\})$ be the weighted sum-rate achieved by the updated orientations \eqref{upeq}. Using the chain rule and the symmetry of $\bm T^{(t)}_{k,m}=\bm I-\bm f^{(t)}_{k,m}(\bm f^{(t)}_{k,m})^{\!T}$, we have

\vspace{-10pt}

{\small 
\begin{align}
   \frac{d\bm f^{(t+1)}_{k,m}(\rho)}{d\rho}\Big|_{\rho=0}
    &= \bm T^{(t)}_{k,m}\,\bm s^{(t)}_{k,m}, \label{d121}\\
    \frac{d\Phi(\rho)}{d\rho}\Big|_{\rho=0} &= \sum_{k,m}\Big\langle \nabla_{\bm f_{k,m}} R\big(\{\bm f_{k,m}^{(t)}\}\big),\, 
          \frac{d\bm f^{(t+1)}_{k,m}(\rho)}{d\rho}\Big|_{\rho=0}\Big\rangle \notag\\
    &= \sum_{k,m}\!\big\langle \bm g^{(t)}_{k,m},\,\bm T^{(t)}_{k,m}\bm s^{(t)}_{k,m}\big\rangle \notag\\
    &= \sum_{k,m}\!\big\langle \bm T^{(t)}_{k,m}\bm g^{(t)}_{k,m},\,\bm s^{(t)}_{k,m}\big\rangle \notag \\
    &= \sigma^{(t)}, \label{d122}
\end{align}
}

\vspace{-5pt}

\noindent where the last equality uses the fact that $\langle \bm T^{(t)}_{k,m}\bm g^{(t)}_{k,m},\,\bm s^{(t)}_{k,m}\rangle=\langle \bm T^{(t)}_{k,m}\bm g^{(t)}_{k,m},\,\bm d^{(t)}_{k,m}\rangle$ since $\langle \bm T^{(t)}_{k,m}\bm g^{(t)}_{k,m},\,\bm f^{(t)}_{k,m}\rangle=0$. Note that by Proposition~1, we have $\sigma^{(t)}\geq 0$.

If $\sigma^{(t)}=0$, we (by tie-breaking) take $\bm s^{(t)}_{k,m}=\bm f^{(t)}_{k,m}$ so that
$\bm d^{(t)}_{k,m}=\bm 0$ and \eqref{eq:update_rule} yields
$\bm f^{(t+1)}_{k,m}=\bm f^{(t)}_{k,m}$; hence
$R(\{\bm f^{(t+1)}\})=R(\{\bm f^{(t)}\})
=R(\{\bm f^{(t)}\})+c_{\mathrm A}\rho^{(t)}\sigma^{(t)}$, which proves that \eqref{eq:armijo_condition} holds with equality.

If $\sigma^{(t)}>0$, applying the first-order Taylor expansion to $\Phi(\rho)$ gives
\[
\Phi(\rho)=\Phi(0)+\rho\,\sigma^{(t)}+r(\rho),
\]
where $r(\rho)=o(\rho)$, for $\rho\rightarrow 0$. Given $c_{\mathrm A}\in(0,1)$, choose $\bar\rho\in(0,1]$ such that
$|r(\rho)|\le (1-c_{\mathrm A})\,\rho\,\sigma^{(t)}$ for all $\rho\in(0,\bar\rho]$.
Then, for any such $\rho$,
\[
\Phi(\rho)\ \ge\ \Phi(0)+\rho\,\sigma^{(t)}-|r(\rho)|
\ \ge\ \Phi(0)+c_{\mathrm A}\,\rho\,\sigma^{(t)},
\]
which is exactly \eqref{eq:armijo_condition}. With backtracking
$\rho^{(t)}=\beta^{j}$, $\beta\in(0,1)$, choose $j$ large enough so that
$\rho^{(t)}\le\bar\rho$, and then the Armijo condition holds, completing the proof.
\end{IEEEproof}

\end{appendix}

\vfill

\end{document}